\documentclass[12pt, onecolumn, draftcls]{IEEEtran}    

\IEEEoverridecommandlockouts                              
\title{Source and Channel Coding for Correlated Sources Over Multiuser Channels}

\author{Deniz G\"{u}nd\"{u}z, Elza Erkip, Andrea Goldsmith, H. Vincent Poor
\thanks{The material in this paper was presented in
part at the 2nd UCSD Information Theory and Applications Workshop (ITA), San Diego, CA, Jan. 2007, at the Data Compression Conference (DCC), Snowbird, UT, March 2007, and at the IEEE International Symposium on Information Theory (ISIT), Nice, France, June 2007.}
\thanks{This research was supported in part by the U.S. National Science Foundation under Grants  ANI-03-38807, CCF-04-30885, CCF-06-35177, CCF-07-28208, and CNS-06-25637 and DARPA ITMANET program under grant 1105741-1-TFIND and the ARO under MURI award W911NF-05-1-0246.}
\thanks{Deniz G\"{u}nd\"{u}z is with the Department of Electrical Engineering, Princeton University, Princeton, NJ, 08544 and with the Department of Electrical Engineering, Stanford University, Stanford, CA, 94305 (email: dgunduz@princeton.edu).}
\thanks{Elza Erkip is with the Department of Electrical and Computer Engineering, Polytechnic Institute of New York University, Brooklyn, NY 11201 (email: elza@poly.edu).}
\thanks{Andrea Goldsmith is with the Department of Electrical Engineering, Stanford University, Stanford, CA, 94305 (email: andrea@systems.stanford.edu).}
\thanks{H. Vincent Poor is with the Department of Electrical Engineering, Princeton University, Princeton, NJ, 08544 (email: poor@princeton.edu).}
}

\date{}
\usepackage[dvips]{hyperref}
\usepackage[dvips]{graphicx}
\usepackage[cmex10]{amsmath}
\usepackage{amsfonts}
\usepackage{amssymb}
\usepackage{amscd}
\usepackage{psfrag}

\newtheorem{thm}{Theorem}[section]
\newtheorem{cor}[thm]{Corollary}
\newtheorem{lem}[thm]{Lemma}
\newtheorem{prop}[thm]{Proposition}
\newtheorem{defn}{Definition}[section]
\newtheorem{rem}{Remark}[section]

\begin{document}
\maketitle
\thispagestyle{empty}

\vspace{-.5in}
\begin{abstract}
Source and channel coding over multiuser channels in which receivers have access to correlated source side information is considered. For several multiuser channel models necessary and sufficient conditions for optimal separation of the source and channel codes are obtained. In particular, the multiple access channel, the compound multiple access channel, the interference channel and the two-way channel with correlated sources and correlated receiver side information are considered, and the optimality of separation is shown to hold for certain source and side information structures. Interestingly, the optimal separate source and channel codes identified for these models are not necessarily the optimal codes for the underlying source coding or the channel coding problems. In other words, while separation of the source and channel codes is optimal, the nature of these optimal codes is impacted by the joint design criterion.
\end{abstract}


\section{Introduction}\label{s:intro}

Shannon's source-channel separation theorem states that, in point-to-point communication systems, a source can be reliably transmitted over a channel if and only if the minimum source coding rate is below the channel capacity \cite{CsiszarKorner}. This means that a simple comparison of the rates of the optimal source and channel codes for the underlying source and channel distributions, respectively, suffices to conclude whether reliable transmission is possible or not. Furthermore, the separation theorem dictates that the source and channel codes can be designed independently without loss of optimality. This theoretical optimality of modularity has reinforced the notion of network layers, leading to the separate development of source and channel coding aspects of a communication system. The separation theorem holds for stationary and ergodic sources and channels under the usual information theoretic assumptions of infinite delay and complexity (see \cite{Verdu_Steinberg} for more general conditions under which separation holds). However, Shannon's source-channel separation theorem does not generalize to multiuser networks.

Suboptimality of separation for multiuser systems was first shown by Shannon in \cite{Shannon2}, where an example of correlated source transmission over the two-way channel was provided. Later, a similar observation was made for transmitting correlated sources over multiple access channels (MACs) in  \cite{Cover_Salehi}. The example provided in \cite{Cover_Salehi} reveals that comparison of the Slepian-Wolf source coding region \cite{SlepianWolf73a} with the capacity region of the underlying MAC is not sufficient to decide whether reliable transmission can be realized.

In general communication networks have multiple sources available at the network nodes, where the source data must be transmitted to its destination in a lossless or lossy fashion. Some (potentially all) of the nodes can transmit while some (potentially all) of the nodes can receive noisy observations of the transmitted signals. The communication channel is characterized by a probability transition matrix from the inputs of the transmitting terminals to the outputs of the receiving terminals. We assume that all the transmissions share a common communications medium; special cases such as orthogonal transmission can be specified through the channel transition matrix. The sources come from an arbitrary joint distribution, that is, they might be correlated. For this general model, the problem we address is to determine whether the sources can be transmitted losslessly or within the required fidelity to their destinations for a given number of channel uses per source sample (cupss), which is defined to be the \emph{source-channel rate} of the joint source channel code. Equivalently, we might want to find the minimum source-channel rate that can be achieved either reliably (for lossless reconstruction) or with the required reconstruction fidelity (for lossy reconstruction).

The problem of jointly optimizing source coding along with the multiuser channel coding in this very general setting is extremely complicated. If the channels are assumed to be noise-free finite capacity links, the problem reduces to a multiterminal source coding problem \cite{CsiszarKorner}; alternatively, if the sources are independent, then we must find the capacity region of a general communication network. Furthermore, considering that we do not have a separation result for source and channel coding even in the case of very simple networks, the hope for solving this problem in the general setting is slight.

Given the difficulty of obtaining a general solution for arbitrary networks, our goal here is to analyze in detail simple, yet fundamental, building blocks of a larger network, such as the multiple access channel, the broadcast channel, the interference channel and the two-way channel. Our focus in this work is on lossless transmission and our goal is to characterize the set of achievable source-channel rates for these canonical networks. Four fundamental questions that need to be addressed for each model can be stated as follows:
\begin{enumerate}
  \item Is it possible to characterize the optimal source-channel rate of the network (i.e., the minimum number of channel uses per source sample (cupss) required for lossless transmission) in a computable way?
  \item Is it possible to achieve the optimum source-channel rate by statistically independent source and channel codes? By statistical independent source and channel codes, we mean that the source and the channel codes are designed solely based on the distributions of the source and the channel distributions, respectively. In general, these codes need not be the optimal codes for the underlying sources or the channel.
  \item Can we determine the optimal source-channel rate by simply comparing the source coding rate region with the capacity region?
  \item If the comparison of these canonical regions is not sufficient to obtain the optimal source-channel rate, can we identify alternative finite dimensional source and channel rate regions pertaining to the source and channel distributions, respectively, whose comparison provides us the necessary and sufficient conditions for the achievability of a source-channel rate?
\end{enumerate}

If the answer to question (3) is affirmative for a given setup, this would maintain the optimality of the layered approach described earlier, and would correspond to the multiuser version of Shannon's source-channel separation theorem. However, even when this classical layered approach is suboptimal, we can still obtain modularity in the system design, if the answer to question (2) is affirmative, in which case the optimal source-channel rate can be achieved by statistically independent source and channel codes, without taking the joint distribution into account. 

In the point-to-point setting, the answer to question (3) is affirmative, that is, the minimum source-channel rate is simply the ratio of the source entropy to the channel capacity; hence these two numbers are all we need to identify the necessary and sufficient conditions for the achievability of a source-channel rate. Therefore, a source code that meets the entropy bound when used with a capacity achieving channel code results in the best source-channel rate. In multiuser scenarios, we need to compare more than two numbers. In classical Shannon separation, it is required that the intersection of the source coding rate region for the given sources and the capacity region of the underlying multiuser channel is not empty. This would definitely lead to modular source and channel code design without sacrificing optimality. However, we show in this work that, in various multiuser scenarios, even if this is not the case for the canonical source coding rate region and the capacity region, it might still be possible to identify alternative finite dimensional rate regions for the sources and the channel, respectively, such that comparison of these rate regions provide the necessary and sufficient conditions for the achievability of a source-channel rate. Hence, the answer to question (4) can be affirmative even if the answer to question (3) is negative. Furthermore, we show that in those cases we also have an affirmative answer to question (2), that is, statistically independent source and channel codes are optimal.

Following \cite{Tuncel}, we will use the following definitions to differentiate between the two types of source-channel separation. \emph{Informational separation} refers to classical separation in the Shannon sense, in which concatenating optimal source and channel codes for the underlying source and channel distributions result in the optimal source-channel coding rate. Equivalently, in informational separation, comparison of the underlying source coding rate region and the channel capacity region is sufficient to find the optimal source-channel rate and the answer to question (3) is affirmative. \emph{Operational separation}, on the other hand, refers to statistically independent source and channel codes that are not necessarily the optimal codes for the underlying source or the channel. Optimality of operational separation allows the comparison of more general source and channel rate regions to provide necessary and sufficient conditions for achievability of a source-channel rate, which suggests an affirmative answer to question (4). These source and channel rate regions are required to be dependent solely on the source and the channel distributions, respectively; however, these regions need not be the canonical source coding rate region or the channel capacity region. Hence, the source and channel codes that achieve different points of these two regions will be statistically independent, providing an affirmative answer to question (2), while individually they may not be the optimal source or channel codes for the underlying source compression and channel coding problems. Note that the class of codes satisfying operational separation is larger than that satisfying informational separation. We should remark here that we are not providing precise mathematical definitions for operational and information separation. Our goal is to point out the limitations of the classical separation approach based on the direct comparison of source coding and channel capacity regions.

This paper provides answers to the four fundamental questions about source-channel coding posed above for some special multiuser networks and source structures. In particular, we consider correlated sources available at multiple transmitters communicating with receivers that have correlated side information. Our contributions can be summarized as follows.
\begin{itemize}
  \item In a multiple access channel we show that informational separation holds if the sources are independent given the receiver side information. This is different from the previous separation results \cite{Han}- \cite{EffrosMedard} in that we show the optimality of separation for an arbitrary multiple access channel under a special source structure. We also prove that the optimality of informational separation continue to hold for independent sources in the presence of correlated side information at the receiver, given which the sources are correlated.

  \item We characterize an achievable source-channel rate for compound multiple access channels with side information, which is shown to be optimal for some special scenarios. In particular, optimality holds either when each user's source is independent from the other source and one of the side information sequences, or when there is no multiple access interference at the receivers. For these cases we argue that operational separation is optimal. We further show the optimality of informational separation when the two sources are independent given the side information common to both receivers. Note that the compound multiple access channel model combines both the multiple access channels with correlated sources and the broadcast channels with correlated side information at the receivers.

  \item For an interference channel with correlated side information, we first define the \emph{strong source-channel interference} conditions, which provide a generalization of the usual strong interference conditions \cite{Costa:IT:87}. Our results show the optimality of operational separation under strong source-channel interference conditions for certain source structures.

  \item We consider a two-way channel with correlated sources. The achievable scheme for compound MAC can also be used as an achievable coding scheme in which the users do not exploit their channel outputs for channel encoding (`restricted encoders'). We generalize Shannon's outer bound for two-way channels to correlated sources.
\end{itemize}

Overall, our results characterize the necessary and sufficient conditions for reliable transmission of correlated sources over various multiuser networks, hence answering question (1) for those scenarios. In these cases, the optimal performance is achieved by statistically independent source and channel codes (by either informational or operational separation), thus promising a level of modularity even when simply concatenating optimal source and channel codes is suboptimal. Hence, for the cases where we provide the optimal source-channel rate, we answer questions (2), (3) and (4) as well.

The remainder of the paper is organized as follows. We review the prior work on joint source-channel coding for multiuser systems in Section \ref{s:prior}, and the notations and the technical tools that will be used throughout the paper in Section \ref{s:prem}. In Section \ref{s:model}, we introduce the system model and the definitions. The next four sections are dedicated to the analysis of special cases of the general system model. In particular, we consider multiple access channel model in Section \ref{s:mac}, compound multiple access channel model in Section \ref{s:cMAC}, interference channel model in Section \ref{s:ic} and finally the two-way channel model in Section \ref{s:twc}. Our conclusions can be found in Section \ref{s:conc} followed by the Appendix.

\section{Prior Work}\label{s:prior}

The existing literature provides limited answers to the four questions stated in Section \ref{s:intro} in specific settings. For the MAC with correlated sources, finite-letter sufficient conditions for achievability of a source-channel rate are given in \cite{Cover_Salehi} in an attempt to resolve the first problem; however, these conditions are later shown not to be necessary by Dueck \cite{Dueck}. The \emph{correlation preserving mapping} technique of \cite{Cover_Salehi} used for achievability is later extended to source coding with side information via multiple access channels in \cite{Ahlswede_Han}, to broadcast channels with correlated sources in \cite{Han_Costa}, and to interference channels in \cite{Kurtas}. In \cite{Pradhan1, Pradhan2} a graph theoretic framework was used to achieve improved source-channel rates for transmitting correlated sources over multiple access and broadcast channels, respectively. A new data processing inequality was proved in \cite{KangUlukus} that is used to derive new necessary conditions for reliable transmission of correlated sources over MACs.

Various special classes of source-channel pairs have been studied in the literature in an effort to resolve the third question above, looking for the most general class of sources for which the comparison of the underlying source coding rate region and the capacity region is sufficient to determine the achievability of a source-channel rate. Optimality of separation in this classical sense is proved for a network of independent, non-interfering channels in \cite{Han}. A special class of the MAC, called the asymmetric MAC, in which one of the sources is available at both encoders, is considered in \cite{Bruyn} and the classical source-channel separation optimality is shown to hold with or without causal perfect feedback at either or both of the transmitters. In \cite{EffrosMedard}, it is shown that for the class of MACs for which the capacity region cannot be enlarged by considering correlated channel inputs, classical separation is optimal. Note that all of these results hold for a special class of MACs and arbitrary source correlations.

There have also been results for joint source-channel codes in broadcast channels. Specifically, in \cite{Tuncel}, Tuncel finds the optimal source-channel rate for broadcasting a common source to multiple receivers having access to different correlated side information sequences, thus answering the first question. This work also shows that the comparison of the broadcast channel capacity region and the minimum source coding rate region is not sufficient to decide whether reliable transmission is possible. Therefore, the classical informational source-channel separation, as stated in the third question, does not hold in this setup. Tuncel also answers the second and fourth questions, and suggests that we can achieve the optimal source-channel rate by source and channel codes that are statistically independent, and that, for the achievability of a source-channel rate $b$, the intersection of two regions, one solely depending on the source distributions, and a second one solely depending on the channel distributions, is necessary and sufficient. The codes proposed in \cite{Tuncel} consist of a source encoder that does not use the correlated side information, and a joint source-channel decoder; hence they are not stand-alone source and channel codes\footnote{Here we note that the joint source-channel decoder proposed by Tuncel in \cite{Tuncel} can also be implemented by separate source and channel decoders in which the channel decoder is a list decoder \cite{Elias} that outputs a list of possible channel inputs. However, by stand-alone source and channel codes, we mean unique decoders that produce a single codeword output, as it is understood in the classical source-channel separation theorem of Shannon.}. Thus the techniques in \cite{Tuncel} require the design of new codes appropriate for joint decoding with the side information; however, it is shown in \cite{Gunduz:ITW:07} that the same performance can be achieved by using separate source and channel codes with a specific message passing mechanism between the source/channel encoders/decoders. Therefore we can use existing near-optimal codes to achieve the theoretical bound.

Broadcast channel in the presence of receiver message side information, i.e., messages at the transmitter known partially or totally at one of the receivers, is also studied from the perspective of achievable rate regions in \cite{Wu:ISIT:07} - \cite{Kang:ISIT:08}. The problem of broadcasting with receiver side information is also encountered in the two-way relay channel problem studied in \cite{Oechtering:IT:08}, \cite{Gunduz:Allerton:08}.

\section{Preliminaries}\label{s:prem}

\subsection{Notation}
In the rest of the paper we adopt the following notational conventions. Random variables will be denoted by capital letters while their realizations will be denoted by the respective lower case letters. The alphabet of a scalar random variable $X$ will be denoted by the corresponding calligraphic letter $\mathcal{X}$, and the alphabet of the $n$-length vectors over the $n$-fold Cartesian product by $\mathcal{X}^n$. The cardinality of set $\mathcal{X}$ will be denoted by $|\mathcal{X}|$. The random vector $(X_1, \ldots, X_n)$ will be denoted by $X^n$ while the vector $(X_i, X_{i+1}, \ldots, X_n)$ by $X_i^n$, and their realizations, respectively, by $(x_1, \ldots, x_n)$ or $x^n$ and $(x_i, x_{i+1}, \ldots, x_n)$ or $x_i^n$.

\subsection{Types and Typical Sequences}
Here, we briefly review the notions of types and strong typicality that will be used in the paper. Given a distribution $p_X$, the type $P_{x^n}$ of an $n$-tuple $x^n$ is the empirical distribution
\[P_{x^n} = \frac{1}{n} N(a|x^n)\]
where $N(a|x^n)$ is the number of occurances  of the letter $a$ in $x^n$. The set of all $n$-tuples $x^n$ with type $Q$ is called the type class $Q$ and denoted by $T_Q^n$. The set of $\delta$-strongly typical $n$-tuples according to $P_X$ is denoted by $T_{[X]_\delta}^n$ and is defined by
\[T_{[X]_\delta}^n = \left\{ x\in \mathcal{X}^n : \left| \frac{1}{n} N(a|x^n)-P_X(a)\right|\leq \delta \forall a \in \mathcal{X} \mbox{ and } N(a|x^n)=0 \mbox{ whenever } P_X(x)=0   \right\}. \]

The definitions of type and strong typicality can be extended to joint and conditional distributions in a similar manner \cite{CsiszarKorner}. The following results concerning typical sets will be used in the sequel. We have
\begin{eqnarray}\label{type0}
\left|\frac{1}{n}\log |T_{[X]_\delta}^n| - H(X) \right| \leq \frac{\delta}{|\mathcal{X}|}
\end{eqnarray}
for sufficiently large $n$. Given a joint distribution $P_{XY}$, if $(x_i, y_i)$ is drawn independent and identically distributed (i.i.d.) with $P_X P_Y$ for $i=1,\ldots,n$, where $P_X$ and $P_Y$ are the marginals, then
\begin{eqnarray}\label{type1}
\mathrm{Pr} \{(x^n, y^n) \in T_{[XY]_\delta}^n \} \leq 2^{-n(I(X;Y)-3\delta)}.
\end{eqnarray}

Finally, for a joint distribution $P_{XYZ}$, if $(x_i, y_i, z_i)$ is drawn i.i.d. with $P_X P_Y P_Z$ for $i=1,\ldots,n$, where $P_X$, $P_Y$ and $P_Z$ are the marginals, then
\begin{eqnarray}\label{type2}
\mathrm{Pr} \{(x^n, y^n, z^n) \in T_{[XYZ]_\delta}^n \} \leq 2^{-n(I(X;Y,Z)+ I(Y;X,Z) + I(Z;Y,X) - 4 \delta)}.
\end{eqnarray}

\section{System Model}\label{s:model}

We introduce the most general system model here. Throughout the paper we consider various special cases, where the restrictions are stated explicitly for each case.

\begin{figure*}
\centering
\psfrag{S1}{$S_1^m$} \psfrag{S2}{$S_2^m$}
\psfrag{X1}{$X_1^n$} \psfrag{X2}{$X_2^n$}
\psfrag{Y1}{$Y_1^n$} \psfrag{Y2}{$Y_2^n$}
\psfrag{W1}{$W_1^m$} \psfrag{W2}{$W_2^m$}
\psfrag{hS1}{$(\hat{S}_{1,1}^m, \hat{S}_{1,2}^m)$} \psfrag{hS2}{$(\hat{S}_{2,1}^m, \hat{S}_{2,2}^m)$}
\psfrag{pxy}{$p(y_1,y_2|x_1,x_2)$}
\psfrag{Tx1}{Transmitter 1} \psfrag{Tx2}{Transmitter 2}
\psfrag{Rx1}{Receiver 1} \psfrag{Rx2}{Receiver 2}
\includegraphics[width=6in]{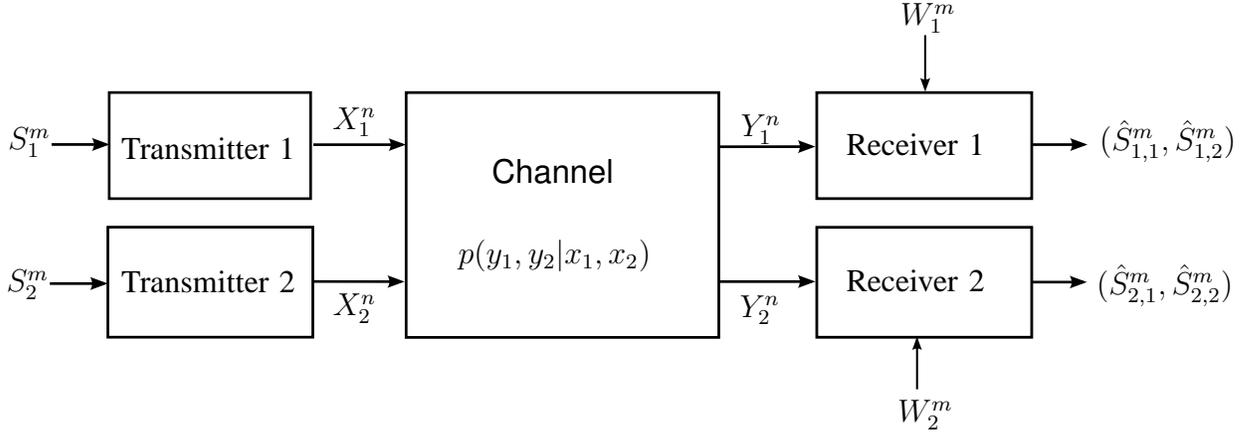}
\caption{The general system model for transmitting correlated sources over multiuser channels with correlated side information. In the MAC scenario, we have only one receiver $\mathrm{Rx}_1$; in the compound MAC scenario, we have two receivers which want to receive both sources, while in the interference channel scenario, we have two receivers, each of which wants to receive only its own source. The compound MAC model reduces to the ``restricted'' two-way channel model when $W_i^m = S_i^m$ for $i=1,2$.
} \label{f:cMAC}
\end{figure*}

We consider a network of two transmitters $\mathrm{Tx}_1$ and $\mathrm{Tx}_2$, and two receivers $\mathrm{Rx}_1$ and $\mathrm{Rx}_2$.  For $i=1,2$, the transmitter  $\mathrm{Tx}_i$ observes the output of a discrete memoryless (DM) source $S_i$, while the receiver $\mathrm{Rx}_i$ observes DM side information $W_i$. We assume that the source and the side information sequences, $\{S_{1,k}, S_{2,k}, W_{1,k}, W_{2,k}\}_{k=1}^\infty$ are i.i.d. and are drawn according to a joint probability mass function (p.m.f.) $p(s_1, s_2, w_1, w_2)$ over a finite alphabet $\mathcal{S}_1 \times  \mathcal{S}_2 \times \mathcal{W}_1 \times \mathcal{W}_2$. The transmitters and the receivers all know this joint p.m.f., but have no direct access to each other's information source or the side information.

The transmitter $\mathrm{Tx}_i$ encodes its source vector $S_i^m=(S_{i,1},\ldots,S_{i,m})$ into a channel codeword $X_i^n=(X_{i,1},\ldots,X_{i,n})$ using the encoding function
\begin{eqnarray}
f_i^{(m,n)}: \mathcal{S}_i^m \rightarrow \mathcal{X}_i^n,
\end{eqnarray}
for $i=1,2$. These codewords are transmitted over a DM channel to the receivers, each of which observes the output vector $Y_i^n=(Y_{i,1},\ldots,Y_{i,n})$. The input and output alphabets $\mathcal{X}_i$ and $\mathcal{Y}_i$ are all finite. The DM channel is characterized by the conditional distribution $P_{Y_1, Y_2|X_1,X_2}(y_1, y_2|x_1,x_2)$.

Each receiver is interested in one or both of the sources depending on the scenario. Let receiver $\mathrm{Rx}_i$ form the estimates of the source vectors $S_1^m$ and $S_2^m$, denoted by $\hat{S}_{i,1}^m$ and $\hat{S}_{i,2}^m$, based on its received signal $Y_i^n$ and the side information vector $W_i^m=(W_{i,1},\ldots,W_{i,m})$ using the decoding function
\begin{eqnarray}
g_i^{(m,n)}: \mathcal{Y}_i^n \times \mathcal{W}_i^m \rightarrow \mathcal{S}_1^m \times \mathcal{S}_2^m.
\end{eqnarray}
Due to the reliable transmission requirement, the reconstruction alphabets are the same as the source alphabets. In the MAC scenario, there is only one receiver $\mathrm{Rx}_1$, which wants to receive both of the sources $S_1$ and $S_2$. In the compound MAC scenario, both receivers want to receive both sources, while in the interference channel scenario, each receiver wants to receive only its own transmitter's source. The two-way channel scenario cannot be obtained as a special case of the above general model, as the received channel output at each user can be used to generate channel inputs. On the other hand, a ``restricted'' two-way channel model, in which the past channel outputs are only used for decoding, is a special case of the above compound channel model with $W_i^m = S_i^m$ for $i=1,2$. Based on the decoding requirements, the error probability of the system, $P_e^{(m,n)}$ will be defined separately for each model. Next, we define the source-channel rate of the system.

\begin{defn}
We say that source-channel rate $b$ is \emph{achievable} if, for every $\epsilon>0$, there exist positive integers $m$ and $n$ with $n/m=b$ for which we have encoders $f_1^{(m,n)}$ and $f_2^{(m,n)}$, and decoders $g_1^{(m,n)}$ and $g_2^{(m,n)}$ with decoder outputs $(\hat{S}_{i,1}^m, \hat{S}_{i,2}^m) = g_i(Y_i^n, W_i^m)$, such that $P_e^{(m,n)} < \epsilon$.
\vspace{.1in}
\end{defn}

\section{Multiple Access Channel}\label{s:mac}

We first consider the multiple access channel, in which we are interested in the reconstruction at receiver $\mathrm{Rx}_1$ only. For encoders $f_i^{(m,n)}$ and a decoder $g_1^{(m,n)}$, the probability of error for the MAC is defined as follows:

\begin{eqnarray}
P_e^{(m,n)} &\triangleq& Pr\{(S_1^m, S_2^m) \neq (\hat{S}_{1,1}^m, \hat{S}_{1,2}^m)\} \nonumber \\
&=& \sum_{(s_1^m,s_2^m) \in \mathcal{S}_1^m \times \mathcal{S}_2^m} p(s_1^m, s_2^m) P\{(\hat{s}_{1,1}^m, \hat{s}_{1,2}^m) \neq (s_1^m, s_2^m) | (S_1^m,S_2^m) = (s_1^m,s_2^m)\}. \nonumber
\end{eqnarray}


Note that this model is more general than that of \cite{Cover_Salehi} as it considers the availability of correlated side information at the receiver \cite{Gunduz:ISIT:07}. We first generalize the achievability scheme of \cite{Cover_Salehi} to our model by using the correlation preserving mapping technique of \cite{Cover_Salehi}, and limiting the source-channel rate $b$ to $1$. Extension to other rates is possible as in Theorem 4 of \cite{Cover_Salehi}.

\begin{thm}\label{t:ach1}
Consider arbitrarily correlated sources $S_1$ and $S_2$ over the DM MAC with receiver side information $W_1$. Source-channel rate $b=1$ is achievable if
\begin{eqnarray}
H(S_1|S_2, W_1) &<&  I(X_1 ; Y_1 | X_2, S_2, W_1, Q), \nonumber \\
H(S_2|S_1, W_1) &<&  I(X_2 ; Y_1 | X_1, S_1, W_1, Q), \nonumber \\
H(S_1, S_2| U, W_1) &<&  I(X_1, X_2 ; Y_1 | U, W_1, Q), \nonumber
\end{eqnarray}
and
\begin{eqnarray}
H(S_1, S_2| W_1) &<&  I(X_1, X_2 ; Y_1 | W_1), \nonumber
\end{eqnarray}
for some joint distribution
\[p(q, s_1, s_2, w_1, x_1, x_2, y_1)=p(q)p(s_1, s_2, w_1)p(x_1|q,s_1) p(x_2|q,s_2) p(y_1|x_1, x_2)\]
and
\[U=f(S_1)=g(S_2)\]
is the common part of $S_1$ and $S_2$ in the sense of G\`{a}cs and K\"{o}rner \cite{GacsKorner}. We can bound the cardinality of $Q$ by $\min\{|\mathcal{X}_1|\cdot |\mathcal{X}_2|, |\mathcal{Y}|\}$.
\vspace{.2in}
\end{thm}

We do not give a proof here as it closely resembles the one in \cite{Cover_Salehi}. Note that correlation among the sources and the side information both condenses the left hand side of the above inequalities, and enlarges their right hand side, compared to transmitting independent sources. While the reduction in entropies on the left hand side is due to Slepian-Wolf source coding, the increase in the right hand side is mainly due to the possibility of generating correlated channel codewords at the transmitters. Applying distributed source coding followed by MAC channel coding, while reducing the redundancy, would also lead to the loss of possible correlation among the channel codewords. However, when $S_1- W_1-S_2$ form a Markov chain, that is, the two sources are independent given the side information at the receiver, the receiver already has access to the correlated part of the sources and it is not clear whether additional channel correlation would help. The following theorem suggests that channel correlation preservation is not necessary in this case and source-channel separation in the informational sense is optimal.

\begin{thm}\label{t:sep_mac}
Consider transmission of arbitrarily correlated sources $S_1$ and $S_2$ over the DM MAC with receiver side information $W_1$, for which the Markov relation $S_1-W_1-S_2$ holds. Informational separation is optimal for this setup, and the source-channel rate $b$ is achievable if
\begin{subequations}\label{e:MAC}
\begin{eqnarray}
H(S_1|W_1) &<& b \cdot I(X_1 ; Y_1 | X_2, Q), \label{e:MAC1} \\
H(S_2|W_1) &<& b \cdot I(X_2 ; Y_1 | X_1, Q), \label{e:MAC2}
\end{eqnarray}
and
\begin{eqnarray}
H(S_1|W_1) + H(S_2|W_1) &<& b \cdot I(X_1, X_2 ; Y_1 |Q), \label{e:MAC3}
\end{eqnarray}
\end{subequations}
for some joint distribution
\begin{eqnarray}\label{jd:MAC}
p(q, x_1, x_2, y_1) = p(q)p(x_1|q)p(x_2|q)p(y_1|x_1, x_2),
\end{eqnarray}
with $|\mathcal{Q}| \leq 4$.

Conversely, if the source-channel rate $b$ is achievable, then the inequalities in (\ref{e:MAC}) hold with $<$ replaced by $\leq$ for some joint distribution of the form given in (\ref{jd:MAC}).
\end{thm}

\begin{proof}
We start with the proof of the direct part. We use Slepian-Wolf source coding followed by multiple access channel coding as the achievability scheme; however, the error probability analysis needs to be outlined carefully since for the rates within the rate region characterized by the right-hand side of (\ref{e:MAC}) we can achieve arbitrarily small \textit{average error probability} rather than the \textit{maximum error probabilit}y \cite{CsiszarKorner}. We briefly outline the code generation and encoding/decoding steps.

Consider a rate pair $(R_1, R_2)$ satisfying
\begin{subequations}\label{e:SW}
\begin{eqnarray}
H(S_1|W_1) < & R_1 & < b \cdot I(X_1 ; Y_1 | X_2, Q), \label {SW1} \\
H(S_2|W_1) < & R_2 & < b \cdot I(X_2 ; Y_1 | X_1, Q) ,\label {SW2}
\end{eqnarray}
and
\begin{eqnarray}
H(S_1|W_1) + H(S_2|W_1) < & R_1 + R_2 & < b \cdot I(X_1, X_2 ; Y_1 |Q). \label {SW3}
\end{eqnarray}
\end{subequations}

\textit{Code generation:} At transmitter $k$, $k=1,2$, independently assign every $s_i^m \in \mathcal{S}_i^m$ to one of the $2^{mR_k}$ bins with uniform distribution. Denote the bin index of $s_k^m$ by $i_k(s_k^m) \in \{1,\ldots, 2^{mR_k}\}$. This constitutes the Slepian-Wolf source code.

Fix $p(q)$, $p(x_1|q)$ and $p(x_2|q)$ such that the conditions in (\ref{e:MAC}) are satisfied. Generate $q^n$ by choosing $q_i$ independently from $p(q)$ for $i=1,\ldots,n$. For each source bin index $i_k = 1,\ldots, 2^{mR_k}$ of transmitter $k$, $k=1,2$, generate a channel codeword $x_k^n(i_k)$ by choosing $x_{ki}(i_k)$ independently from $p(x_k|q_i)$. This constitutes the MAC code.

\textit{Encoders:} We use the above separate source and the channel codes for encoding. The source encoder $k$ finds the bin index of $s_k^m$ using the Slepian-Wolf source code, and forwards it to the channel encoder. The channel encoder transmits the codeword $x_k^n$ corresponding to the source bin index using the MAC code.

\textit{Decoder:} We use separate source and channel decoders. Upon receiving $y_1^n$, the channel decoder tries to find the indices $(i_1', i_2')$ such that the corresponding channel codewords satisfy $(q^n, x_1^n(i_1'), x_2^n(i_2')) \in T^n_{[QX_1X_2Y]_\delta}$. If one such pair is found, call it $(i_1',i_2')$. If no or more than one such pair is found, declare an error.

Then these indices are provided to the source decoder. Source decoder tries to find $\hat{s}_{1,k}^m$ such that $i_k(\hat{s}_k^m) = i_k'$ and $(\hat{s}_k^m, W_1^m) \in T^m_{[S_kW_1]_\delta}$. If one such pair is found, it is declared as the output. Otherwise, an error is declared.

\textit{Probability of error analysis:} For brevity of the expressions, we define $\mathbf{s} = (s_1^m,s_2^m)$, $\mathbf{S} = (S_1^m,S_2^m)$ and $\hat{\mathbf{s}} = (\hat{s}_{1,1}^m, \hat{s}_{1,2}^m)$. The indices corresponding to the sources are denoted by $\mathbf{i} = (i_1(s_1^m), i_2(s_2^m))$, and the indices estimated at the channel decoder are denoted by $\mathbf{i}' = (i_1',i_2')$. The average probability of error can be written as follows:
\begin{eqnarray}
P_e^{(m,n)} &\triangleq& \sum_{\mathbf{s}} P\{\hat{\mathbf{s}} \neq \mathbf{s} | \mathbf{S} = \mathbf{s}\}  p(\mathbf{s}) \nonumber \\
    &= & \sum_{\mathbf{s}} \left[ P\{\hat{\mathbf{s}} \neq \mathbf{s} | \mathbf{i}=\mathbf{i}', \mathbf{S} = \mathbf{s}\}  p(\mathbf{i}=\mathbf{i}' | \mathbf{S} = \mathbf{s}) +  P\{\hat{\mathbf{s}} \neq \mathbf{s} | \mathbf{i} \neq \mathbf{i}', \mathbf{S} = \mathbf{s}\}  p(\mathbf{i} \neq \mathbf{i}' | \mathbf{S} = \mathbf{s}) \right] p(\mathbf{s}) \nonumber \\
    &\leq & \sum_{\mathbf{s}} \left[ P\{\hat{\mathbf{s}} \neq \mathbf{s} | \mathbf{i}=\mathbf{i}', \mathbf{S} = \mathbf{s}\} +  p(\mathbf{i} \neq \mathbf{i}' | \mathbf{S} = \mathbf{s}) \right] p(\mathbf{s}) \nonumber \\
    &= & \sum_{\mathbf{s}} P\{\hat{\mathbf{s}} \neq \mathbf{s} | \mathbf{i}=\mathbf{i}', \mathbf{S} = \mathbf{s}\}  p(\mathbf{s}) + \sum_{\mathbf{s}} p(\mathbf{i} \neq \mathbf{i}' | \mathbf{S} = \mathbf{s}) p(\mathbf{s}) \label{e:er:e4}
\end{eqnarray}

Now, in (\ref{e:er:e4}) the first summation is the average error probability given the fact that the receiver knows the indices correctly. This can be made arbitrarily small with increasing $m$, which follows from the Slepian-Wolf theorem. The second term in (\ref{e:er:e4}) is the average error probability for the indices averaged over all source pairs. This can also be written as
\begin{align}
\sum_{\mathbf{s}} p(\mathbf{i} \neq \mathbf{i}' | \mathbf{S} = \mathbf{s}) p(\mathbf{s}) &= \sum_{\mathbf{i}} p(\mathbf{i} \neq \mathbf{i}', \mathbf{I} = \mathbf{i}) \nonumber\\
 & = \sum_{\mathbf{i}} p(\mathbf{i} \neq \mathbf{i}' | \mathbf{I} = \mathbf{i}) p( \mathbf{I} = \mathbf{i}) \nonumber\\
 & = \frac{1}{2^{m(R_1+R_2)}} \sum_{\mathbf{i}} p(\mathbf{i} \neq \mathbf{i}' | \mathbf{I} = \mathbf{i}) \label{e:er:e5}
\end{align}
where (\ref{e:er:e5}) follows from the uniform assignment of the bin indices in the creation of the source code. Note that (\ref{e:er:e5}) is the average error probability expression for the MAC code, and we know that it can also be made arbitrarily small with increasing $m$ and $n$ under the conditions of the theorem \cite{CsiszarKorner}.

We note here that for $b=1$ the direct part can also be obtained from Theorem \ref{t:ach1}. For this, we ignore the common part of the sources and choose the channel inputs independent of the source distributions, that is, we choose a joint distribution of the form
\[p(q, s_1, s_2, w_1, x_1, x_2, y_1)=p(q)p(s_1, s_2, w_1)p(x_1|q) p(x_2|q) p(y_1|x_1, x_2).\]
From the conditional independence of the sources given the receiver side information, both the left and the right hand sides of the conditions in Theorem \ref{t:ach1} can be simplified to the sufficiency conditions of Theorem \ref{t:sep_mac}.

We next prove the converse. We assume $P_e^{(m,n)} \rightarrow 0$ for a sequence of encoders $f_i^{(m,n)}$ $(i=1,2)$ and decoders $g^{(m,n)}$ as $n,m \rightarrow \infty$ with a fixed rate $b=n/m$. We will use Fano's inequality, which states
\begin{eqnarray}
H(S_1^m,S_2^m|\hat{S}_{1,1}^m, \hat{S}_{1,2}^m) &\leq& 1 + m P_e^{(m,n)} \log |\mathcal{S}_1 \times \mathcal{S}_2|, \nonumber \\ \label {fano1}
&\triangleq& m \delta (P_e^{(m,n)}),
\end{eqnarray}
where $\delta (x)$ is a non-negative function that approaches zero as $x \rightarrow 0$. We also obtain
\begin{eqnarray}
H(S_1^m,S_2^m|\hat{S}_{1,1}^m, \hat{S}_{1,2}^m) &\geq& H(S_1^m |\hat{S}_{1,1}^m, \hat{S}_{1,2}^m),\\
&\geq& H(S_1^m |Y_1^n, W_1^m),
\end{eqnarray}
where the first inequality follows from the chain rule of entropy and the nonnegativity of the entropy function for discrete sources, and the second inequality follows from the data processing inequality. Then we have, for $i=1,2$,
\begin{eqnarray} \label {fano2}
H(S_i^m|Y_1^n, W_1^m) \leq m \delta (P_e^{(m,n)}).
\end{eqnarray}

We have
\begin{eqnarray} \label{ineq12}
\frac{1}{n}I(X_1^n;Y_1^n | X_2^n, W_1^m) &\geq& \frac{1}{n} I(S_1^m ;Y_1^n
| W_1^m, X_2^n), \\  \label{ineq14}
        &=& \frac{1}{n} [H(S_1^m| W_1^m ,X_2^n)- H(S_1^m|Y_1^n, W_1^m, X_2^n)], \\  \label{ineq15}
        &=& \frac{1}{n} [H(S_1^m| W_1^m)- H(S_1^m|Y_1^n, W_1^m, X_2^n)], \\  \label{ineq16}
        &\geq& \frac{1}{n} [H(S_1^m| W_1^m)- H(S_1^m|Y_1^n, W_1^m)], \\  \label{ineq17}
        &\geq& \frac{1}{b} \left[ H(S_1| W_1)- \delta (P_e^{(m,n)}) \right], \label{ineq19}
\end{eqnarray}
where $(\ref{ineq12})$ follows from the Markov relation
$S_1^m-X_1^n-Y_1^n$ given $(X_2^n, W_1^m)$; $(\ref{ineq15})$ from the
Markov relation $X_2^n-W_1^m-S_1^m$; $(\ref{ineq16})$ from the fact
that conditioning reduces entropy; and $(\ref{ineq17})$ from the
memoryless source assumption and from (\ref{fano1}) which uses
Fano's inequality.

On the other hand, we also have
\begin{eqnarray}\label{eqn1aa_1}
 I(X_1^n; Y_1^n | X_2^n, W_1^m) &=& H(Y_1^n|X_2^n, W_1^m) - H(Y_1^n|X_1^n, X_2^n, W_1^m), \\ \label{eqn1a_1}
 &=& H(Y_1^n|X_2^n, W_1^m) - \sum_{i=1}^n H(Y_{1,i}|Y_1^{i-1}, X_1^n, X_2^n, W_1^m) , \\ \label{eqn1a_2}
 &=& H(Y_1^n|X_2^n, W_1^m) - \sum_{i=1}^n H(Y_{1,i}| X_{1i}, X_{2i}, W_1^m),   \\ \label{eqn1a_3}
 &\leq& \sum_{i=1}^n H(Y_{1,i}|X_{2i}, W_1^m) - \sum_{i=1}^n H(Y_{1,i}| X_{1i}, X_{2i}, W_1^m),    \\ \label{eqn1a_5}
 &=& \sum_{i=1}^n I(X_{1i}; Y_{1,i}|X_{2i}, W_1^m),   \label{eqn1a_6}
\end{eqnarray}
where (\ref{eqn1a_1}) follows from the chain rule; (\ref{eqn1a_2})
from the memoryless channel assumption; and (\ref{eqn1a_3}) from the
chain rule and the fact that conditioning reduces entropy.

For the joint mutual information we can write the following set of inequalities:
\begin{eqnarray} \label{ineq21}
\frac{1}{n}I(X_1^n, X_2^n;Y_1^n |W_1^m) &\geq& \frac{1}{n} I(S_1^m,S_2^m
;Y_1^n | W_1^m), \\  \label{ineq24}
        &=& \frac{1}{n} [H(S_1^m,S_2^m| W_1^m)- H(S_1^m,S_2^m|Y_1^n, W_1^m)], \\  \label{ineq25}
        &=& \frac{1}{n} [H(S_1^m| W_1^m) + H(S_2^m| W_1^m) - H(S_1^m,S_2^m|Y_1^n, W_1^m)], \\  \label{ineq26}
        &\geq& \frac{1}{n} [H(S_1^m| W_1^m) + H(S_2^m| W_1^m) - H(S_1^m,S_2^m|\hat{S}_1^m,\hat{S}_2^m), \\  \label{ineq27}
        &\geq& \frac{1}{b} \bigg[H(S_1| W_1) + H(S_2| W_1) - \delta (P_e^{(m,n)}) \bigg],
\end{eqnarray}
where $(\ref{ineq21})$ follows from the Markov relation $(S_1^m,
S_2^m)- (X_1^n, X_2^n) -Y_1^n$ given $W_1^m$; $(\ref{ineq25})$ from the
Markov relation $S_2^m-W_1^m-S_1^m$; $(\ref{ineq26})$ from the fact
that $(S_1^m,S_2^m)- (Y_1^n, W_1^m)-(\hat{S}_1^m, \hat{S}_2^m)$ form a Markov chain; and $(\ref{ineq27})$ from the memoryless source assumption and
from (\ref{fano1}) which uses Fano's inequality.

By following similar arguments as in (\ref{eqn1aa_1})-(\ref{eqn1a_6}) above, we can also show that
\begin{eqnarray}\label{eqn_2aa}
 I(X_1^n, X_2^n; Y_1^n |W_1^m) &\leq& \sum_{i=1}^n I(X_{1i},X_{2i} ; Y_{1,i} |W_1^m). \label{eqn3a_6}
\end{eqnarray}

Now, we introduce a time-sharing random variable $\bar{Q}$ independent of all other random variables. We have $\bar{Q}=i$ with probability $1/n$, $i \in \{1,2,\ldots,n\}$. Then we can write
\begin{eqnarray}
\frac{1}{n}  I(X_1^n; Y_1^n | X_2^n, W_1^m) &\leq& \frac{1}{n} \sum_{i=1}^n I(X_{1i}; Y_{1,i}|X_{2i}, W_1^m), \\
 &=& \frac{1}{n} \sum_{i=1}^n I(X_{1\bar{q}}; Y_{\bar{q}}|X_{2\bar{q}}, W_1^m, \bar{Q}=i), \\
 &=& I(X_{1\bar{Q}}; Y_{\bar{Q}}|X_{2\bar{Q}}, W_1^m, \bar{Q}), \\
&=& I(X_1; Y|X_2, Q),
\end{eqnarray}
where $X_1 \triangleq X_{1\bar{Q}}$, $X_2 \triangleq X_{2\bar{Q}}$,
$Y \triangleq Y_{\bar{Q}}$, and $Q \triangleq (W_1^m, \bar{Q})$. Since $S_1^m$ and $S_2^m$, and therefore $X_{1i}$ and $X_{2i}$, are
independent given $W_1^m$, for $q=(w_1^m,i)$ we have
\begin{eqnarray}
Pr\{X_{1}=x_1, X_{2}=x_2 |Q = q\} &=& Pr\{X_{1i}=x_1, X_{2i}=x_2 |W_1^m = w_1^m, \bar{Q}=i\} \nonumber \\
&=& Pr\{X_{1i}=x_1|W_1^m = w_1^m, \bar{Q}=i \} Pr\{X_{2i}=x_2|W_1^m = w_1^m, \bar{Q}=i\} \nonumber \\
&=& Pr\{X_1|Q=q \} \cdot Pr\{X_2|Q=q\}. \nonumber
\end{eqnarray}
Hence, the probability distribution is of the form given in Theorem \ref{t:sep_mac}.

On combining the inequalities above we can obtain
\begin{eqnarray}
H(S_1|W_1)- \delta(P_e^{(m,n)}) \leq b I(X_1; Y|X_2, Q), \label{cond_rate1} \\
H(S_2|W_1)- \delta(P_e^{(m,n)}) \leq b I(X_2; Y|X_1, Q), \label{cond_rate2}
\end{eqnarray}
and
\begin{equation}
H(S_1|W_1) + H(S_2|W_1) - \delta(P_e^{(m,n)}) \leq b I(X_1, X_2; Y|Q). \label{cond_rate3}
\end{equation}

Finally, taking the limit as $m,n \rightarrow \infty$ and letting
$P_e^{(m,n)} \rightarrow  0$ leads to the conditions of the theorem.
\end{proof}

%

To the best of our knowledge, this result constitutes the first example in which the underlying source structure leads to the optimality of (informational) source-channel separation independent of the channel. We can also interpret this result as follows: The side information provided to the receiver satisfies a special Markov chain condition, which enables the optimality of informational source-channel separation. We can also observe from Theorem \ref{t:sep_mac} that the optimal source-channel rate in this setup is determined by identifying the smallest scaling factor $b$ of the MAC capacity region such that the point $(H(S_1|W_1), H(S_2, W_1))$ falls into the scaled region. This answers question (3) affirmatively in this setup.

A natural question to ask at this point is whether providing some side information to the receiver can break the optimality of source-channel separation in the case of independent messages. In the next theorem, we show that this is not the case, and the optimality of informational separation continues to hold.

\begin{thm}\label{t:sep_mac_vc}
Consider independent sources $S_1$ and $S_2$ to be transmitted over the DM MAC with correlated receiver side information $W_1$. If the joint distribution satisfies $p(s_1, s_2, w_1)= p(s_1)p(s_2)p(w_1|s_1, s_2)$, then the source-channel rate $b$ is achievable if
\begin{eqnarray}
H(S_1|S_2, W_1) &<& b \cdot I(X_1 ; Y_1 | X_2, Q), \label{e:MAC1_vc} \\
H(S_2|S_1, W_1) &<& b \cdot I(X_2 ; Y_1 | X_1, Q), \label{e:MAC2_vc}
\end{eqnarray}
and
\begin{eqnarray}
H(S_1, S_2|W_1) &<& b \cdot I(X_1, X_2 ; Y_1 |Q), \label{e:MAC3_vc}
\end{eqnarray}
for some input distribution
\begin{eqnarray}\label{pd_MAC_v2}
p(q, x_1, x_2, y_1) = p(q)p(x_1|q)p(x_2|q)p(y_1|x_1, x_2),
\end{eqnarray}
with $|\mathcal{Q}| \leq 4$.

Conversely, if the source-channel rate $b$ is achievable, then (\ref{e:MAC1_vc})-(\ref{e:MAC3_vc}) hold with $<$ replaced by $\leq$ for some joint distribution of the form given in (\ref{pd_MAC_v2}). Informational separation is optimal for this setup.
\end{thm}
\begin{proof}
The proof is given in Appendix \ref{App:sep_mac_vc}.
\end{proof}

\psfrag{H(SS)}{$\scriptstyle H(S_1|S_2)$} \psfrag{H(SS2)}{$\scriptstyle H(S_2|S_1)$}
\psfrag{H(SW)}{$\scriptstyle H(S_1|W_1)$} \psfrag{H(SW2)}{$\scriptstyle H(S_2|W_1)$}
\psfrag{(ab)}{$\scriptstyle (0.46,0.46)$}
\psfrag{h}{$\scriptstyle 0.5$} \psfrag{1}{$\scriptstyle 1$}
\psfrag{1.5}{$\scriptstyle 1.5$} \psfrag{1.58}{$\scriptstyle 1.58$}
\begin{figure}
\centering
\includegraphics[width=5in]{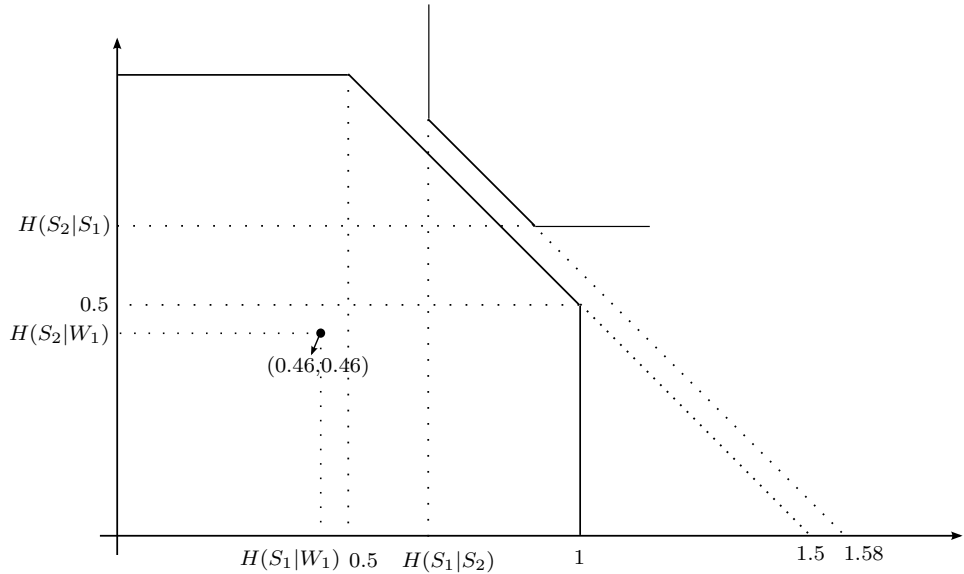}
\caption{Capacity region of the binary adder MAC and the source coding rate regions in the example.} \label{f:regions}
\end{figure}

Next, we illustrate the results of this section with some examples. Consider binary sources and side information, i.e., $\mathcal{S}_1=\mathcal{S}_2=\mathcal{W}_1=\{1,2\}$, with the following joint distribution:
\begin{align}
 P_{S_1S_2W_1}\{S_1=0, S_2=0, W_1=0\} = P_{S_1S_2W_1}\{S_1=1, S_2=1, W_1=1\}  = 1/3 \nonumber
\end{align}
and
\begin{align}
 P_{S_1S_2W_1}\{S_1=0, S_2=1, W_1=0\} = P_{S_1S_2W_1}\{S_1=0, S_2=1, W_1=1\}  = 1/6. \nonumber
\end{align}

As the underlying multiple access channel, we consider a binary input adder channel, in which $\mathcal{X}_1=\mathcal{X}_2=\{0,1\}$, $\mathcal{Y}=\{0,1,2\}$ and \[Y=X_1 +X_2.\] Note that, when the side information $W_1$ is not available at the receiver, this model is the same as the example considered in \cite{Cover_Salehi}, which was used to show the suboptimality of separate source and channel codes over the MAC.

When the receiver does not have access to side information $W_1$, we can identify the separate source and channel coding rate regions using the conditional entropies. These regions are shown in Fig. \ref{f:regions}. The minimum source-channel rate is found as $b=1.58/1.5=1.05$ cupss in the case of separate source and channel codes. On the other hand, it is easy to see that uncoded transmission is optimal in this setup which requires a source-channel rate of $b=1$ cupss. Now, if we consider the availability of the side information $W_1$ at the receiver, we have $H(S_1|W_1)=H(S_2|W_1)=0.46$. In this case, using Theorem \ref{t:sep_mac}, the minimum required source-channel rate is found to be $b=0.92$ cupss, which is lower than the one achieved by uncoded transmission.

Theorem \ref{t:sep_mac_vc} states that, if the two sources are independent, informational source-channel separation is optimal even if the receiver has side information given which independence of the sources no longer holds. Consider, for example, the same binary adder channel in our example. We now consider two independent binary sources with uniform distribution, i.e., $P(S_1=0)=P(S_2=0)=1/2$. Assume that the side information at the receiver is now given by $W_1 =X_1\oplus X_2$, where $\oplus$ denotes the binary xor operation. For these sources and the channel, the minimum source-channel rate without the side information at the receiver is found as $b=1.33$ cupss. When $W_1$ is available at the receiver, the minimum required source-channel rate reduces to $b=0.67$ cupss, which can still be achieved by separate source and channel coding.

Next, we consider the case when the receiver side information is also provided to the transmitters. From the source coding perspective, i.e., when the underlying MAC is composed of orthogonal finite capacity links, it is known that having the side information at the transmitters would not help. However, it is not clear in general, from the source-channel rate perspective, whether providing the receiver side information to the transmitters would improve the performance.

If $S_1-W_1-S_2$ form a Markov chain, it is easy to see that the results in Theorem \ref{t:sep_mac} continue to hold even when $W_1$ is provided to the transmitters. Let $\tilde{S}_i=(S_i, W_1)$ be the new sources for which $\tilde{S}_1-W_1-\tilde{S}_2$ holds. Then, we have the same necessary and sufficient conditions as before, hence providing the receiver side information to the transmitters would not help in this setup.

Now, let $S_1$ and $S_2$ be two independent binary random variables, and $W_1 = S_1 \oplus S_2$. In this setup, providing the receiver side information $W_1$ to the transmitters means that the transmitters can learn each other's source, and hence can fully cooperate to transmit both sources. In this case, source-channel rate $b$ is achievable if
\begin{eqnarray}\label{sideInfoatTx}
H(S_1, S_2|W_1) < b I(X_1, X_2; Y_1)
\end{eqnarray}
for some input distribution $p(x_1, x_2)$, and if source-channel rate $b$ is achievable then (\ref{sideInfoatTx}) holds with $\leq$ for some $p(x_1, x_2)$. On the other hand, if $W_1$ is not available at the transmitters, we can find from Theorem \ref{t:sep_mac_vc} that the input distribution in (\ref{sideInfoatTx}) can only be $p(x_1) p(x_2)$. Thus, in this setup, providing receiver side information to the transmitters potentially leads to a smaller source-channel rate as this additional information may enable cooperation over the MAC, which is not possible without the side information. In our example of independent binary sources, the total transmission rate that can be achieved by total cooperation of the transmitters is $1.58$ bits per channel use. Hence, the minimum source-channel rate that can be achieved when the side information $W_1$ is available at both the transmitters and the receiver is found to be $0.63$ cupss. This is lower than $0.67$ cupps that can be achieved when the side information is only available at the receiver.

We conclude that, as opposed to the pure lossless source coding scenario, having side information at the transmitters might improve the achievable source-channel rate in multiuser systems.

\section{Compound MAC with correlated sources}\label{s:cMAC}

Next, we consider a compound multiple access channel, in which two transmitters wish to transmit their correlated sources reliably to two receivers simultaneously \cite{Gunduz:ISIT:07}. The error probability of this system is given as follows:

\begin{eqnarray}
P_{e}^{(m,n)} &\triangleq& Pr\left\{\bigcup_{k=1,2} (S_1^m, S_2^m) \neq (\hat{S}_{k,1}^m, \hat{S}_{k,2}^m)\right\} \nonumber \\
&=& \sum_{(s_1^m,s_2^m) \in \mathcal{S}_1^m \times \mathcal{S}_2^m} p(s_1^m, s_2^m) P\left\{ \bigcup_{k=1,2} (\hat{s}_{k,1}^m, \hat{s}_{k,2}^m) \neq (s_1^m, s_2^m) \big| (S_1^m,S_2^m) = (s_1^m,s_2^m)\right\}. \nonumber
\end{eqnarray}

The capacity region of the compound MAC is shown to be the intersection of the two MAC capacity regions in \cite{Ahlswede} in the case of independent sources and no receiver side information. However, necessary and sufficient conditions for lossless transmission in the case of correlated sources are not known in general. Note that, when there is side information at the receivers, finding the achievable source-channel rate for the compound MAC is not a simple extension of the capacity region in the case of independent sources. Due to different side information at the receivers, each transmitter should send a different part of its source to different receivers. Hence, in this case we can consider the compound MAC both as a combination of two MACs, and as a combination of two broadcast channels. We remark here that even in the case of single source broadcasting with receiver side information, informational separation is not optimal, but the optimal source-channel rate can be achieved by operational separation as is shown in \cite{Tuncel}.

We first state an achievability result for rate $b=1$, which extends the achievability scheme proposed in \cite{Cover_Salehi} to the compound MAC with correlated side information. The extension to other rates is possible by considering blocks of sources and channels as superletters similar to Theorem 4 in \cite{Cover_Salehi}.

\begin{thm}\label{t:ach_cMAC}
Consider lossless transmission of arbitrarily correlated sources $(S_1, S_2)$ over a DM compound MAC with side information $(W_1, W_2)$ at the receivers as in Fig. \ref{f:cMAC}. Source-channel rate $1$ is achievable if, for $k=1,2$,
\begin{eqnarray}
H(S_1|S_2, W_k) &<&   I(X_1 ; Y_k | X_2, S_2, W_k, Q), \nonumber \\
H(S_2|S_1, W_k) &<&   I(X_2 ; Y_k | X_1, S_1, W_k, Q), \nonumber \\
H(S_1, S_2| U, W_k) &<&  I(X_1, X_2 ; Y_k | U, W_k, Q), \nonumber
\end{eqnarray}
and
\begin{eqnarray}
H(S_1, S_2| W_k) &<&  I(X_1, X_2 ; Y_k | W_k), \nonumber
\end{eqnarray}
for some joint distribution of the form
\[p(q, s_1, s_2, w_1, w_2, x_1, x_2, y_1, y_2) = p(q) p(s_1, s_2, w_1, w_2)p(x_1| q,s_1)p(x_2|q, s_2)p(y_1,y_2|x_1,x_2)\]
and
\[U = f(S_1) = g(S_2)\]
is the common part of $S_1$ and $S_2$ in the sense of G\`{a}cs and K\"{o}rner.
\end{thm}

\begin{proof}
The proof follows by using the correlation preserving mapping scheme of \cite{Cover_Salehi}, and is thus omitted for the sake of brevity.
\end{proof}

In the next theorem, we provide sufficient conditions for the achievability of a source-channel rate $b$. The achievability scheme is based on operational separation where the source and the channel codebooks are generated independently of each other. In particular, the typical source outputs are matched to the channel inputs without any explicit binning at the encoders. At the receiver, a joint source-channel decoder is used, which can be considered as a concatenation of a list decoder as the channel decoder, and a source decoder that searches among the list for the source codeword that is also jointly typical with the side information. However, there are no explicit source and channel codes that can be independently used either for compressing the sources or for independent data transmission over the underlying compound MAC. An alternative coding scheme composed of explicit source and channel coders that interact with each other is proposed in \cite{Gunduz:ITW:07}. However, the channel code in this latter scheme is not the channel code for the underlying multiuser channel either.

\begin{thm}\label{t:ach2}
Consider lossless transmission of arbitrarily correlated sources $S_1$ and $S_2$ over a DM compound MAC with side information $W_1$ and $W_2$ at
the receivers. Source-channel rate $b$ is achievable if, for $k=1,2$,
\begin{eqnarray}
H(S_1|S_2, W_k) < b I(X_1;Y_k |X_2,Q), \label{e:cmac1} \\
H(S_2|S_1, W_k) < b I(X_2;Y_k |X_1,Q), \label{e:cmac2}
\end{eqnarray}
and
\begin{eqnarray}
H(S_1, S_2 |W_k) < b I(X_1, X_2 ;Y_k |Q), \label{e:cmac3}
\end{eqnarray}
for some $|\mathcal{Q}|\leq 4$ and input distribution of the form $p(q, x_1, x_2) = p(q)$ $p(x_1|q)p(x_2|q)$.
\end{thm}

\begin{rem}
The achievability part of Theorem \ref{t:ach2} can be obtained from the achievability of Theorem \ref{t:ach_cMAC}. Here, we constrain the channel input distributions to be independent of the source distributions as opposed to the conditional distribution used in Theorem \ref{t:ach_cMAC}. We provide the proof of the achievability of Theorem \ref{t:ach2} below to illustrate the nature of the operational separation scheme that is used.
\end{rem}

\begin{proof}
Fix $\delta_k>0$ and $\gamma_{k}>0$ for $k=1,2$, and $P_{X_1}$ and $P_{X_2}$. For $b=n/m$ and $k=1,2$, at transmitter
$k$, we generate $M_k=2^{m[H(S_k)+\epsilon/2]}$ i.i.d. length-$m$
source codewords and i.i.d. length-$n$ channel codewords using
probability distributions $P_{S_k}$ and $P_{X_k}$, respectively.
These codewords are indexed and revealed to the receivers as well, and are denoted
by $s_k^m(i)$ and $x_k^n(i)$ for $1\leq i \leq M_k$.

\emph{Encoder:} Each source outcome is directly mapped to a channel
codeword as follows: Given a source outcome $S_k^m$ at transmitter
$m$, we find the smallest $i_k$ such that $S_k^m = s_k^m(i_k)$, and
transmit the codeword $x_k^n(i_k)$. An error occurs if no such $i_k$
is found at either of the transmitters $k=1,2$.

\emph{Decoder:} At receiver $k$, we find the unique pair $(i_1^*, i_2^*)$ that simultaneously satisfies
\begin{eqnarray}
(x_1^n(i_1^*),x_2^n(i_2^*), Y_k^n ) \in \mathcal{T}^{(n)}_{[X_1 X_2 Y]_{\delta_k}}, \nonumber
\end{eqnarray}
and
\begin{eqnarray}
(s_1^m(i_1^*), s_2^m(i_2^*), W_k^m) \in \mathcal{T}^{(m)}_{[S_1 S_2 W_k]_{\gamma_{k}}}, \nonumber
\end{eqnarray}
where $T^{(n)}_{[X]_{\delta}}$ is the set of weakly $\delta$-typical sequences. An error is declared if the $(i_1^*, i_2^*)$ pair is not uniquely determined.

\emph{Probability of error:} We define the following events:
\begin{align}
E_1^k &= \{ S_k^m \neq s_k^m(i), \forall i\}  \nonumber \\
E_2^k &= \{ (s_1^m(i_1), s_2^m(i_2), W_k^m) \notin  T^{(m)}_{[S_1 S_2 W_k]_{\gamma_{k}}} \} \nonumber\\
E_3^k &=  \{ (x_1^n(i_1), x_2^n(i_2), Y_k^n) \notin  T^{(n)}_{[X_1 X_2 Y]_{\delta_k}} \}  \nonumber
\end{align}
and
\begin{align}
E_4^k (j_1,j_2) &=  \{(s_1^m(j_1), s_2^m(j_2), W_k^m) \in  T^{(m)}_{[S_1 S_2 W_k]_{\gamma_{k}}} \mbox{ and } (x_1^n(j_1), x_2^n(j_2), Y_k^n) \in  T^{(n)}_{[X_1 X_2 Y]_{\delta_k}} \}  \nonumber
\end{align}

Here, $E_1$ denotes the error event in which either of the encoders
fails to find a unique source codeword in its codebook that
corresponds to its current source outcome. When such a codeword can
be found, $E_2^k$ denotes the error event in which the sources $S_1^m$ and $S_2^m$ and the side information $W_{k}$ at receiver $k$ are not
jointly typical. On the other hand, $E_3^k$ denotes the
error event in which channel codewords that match the current source
realizations are not jointly typical with the channel output at
receiver $k$. Finally $E_4^k(j_1,j_2)$ is the event that the source codewords corresponding to the indices $j_1$ and $j_2$ are jointly typical with the side information $W_k$ and simultaneously that the channel codewords corresponding to the indices $j_1$ and $j_2$ are jointly typical with the channel output $Y_k$.

Define $P_k^{(m,n)} \triangleq \mathrm{Pr} \{(S_1^m, S_2^m) \neq (\hat{S}_{k,1}^m, \hat{S}_{k,2}^m)\}$. Then $P_e^{(m,n)} \leq \sum_{k=1,2} P_k^{(m,n)}$. Again, from the union bound, we have
\begin{align}\label{err_prob}
P_k^{(m,n)}  \leq & \mathrm{Pr}\{E_1^k\} + \mathrm{Pr}\{E_2^k\} + \mathrm{Pr}\{E_3^k\} + \sum_{\substack{j_1 \neq i_1, \\ j_2 = i_2}} E_4^k(j_1,j_2) &+ \sum_{\substack{j_1 = i_1, \\ j_2 \neq i_2}} E_4^k(j_1,j_2) + \sum_{\substack{j_1 \neq i_1, \\ j_2 \neq i_2}} E_4^k(j_1,j_2),
\end{align}
where $i_1$ and $i_2$ are the correct indices. We have
\begin{align}
E_4^k(j_1,j_2) = \mathrm{Pr} \left\{(s_1^m(j_1), s_2^m(j_2), W_k^m) \in  T^{(m)}_{[S_1, S_2, W_k]_{\gamma_{k}}}\right\} \mathrm{Pr} \left\{ (x_1^n(j_1), x_2^n(j_2), Y_k^n) \in  T^{(n)}_{[X_1, X_2, Y_k]_{\delta_k}} \right\}.
\end{align}
In \cite{Tuncel} it is shown that, for any $\lambda > 0$ and sufficiently large $m$,
\begin{align}
\mathrm{Pr}\{E_1^k\} & = (1-\mathrm{Pr}\{S_k^m = s_k^m(1)\})^{M_k} \nonumber \\
    &\leq \exp^{-2^{-n[H(S_k)+6\lambda]}M_k} \nonumber \\
    & = \exp^{-2^{n[\frac{\epsilon}{2}-6\lambda]}}.
\end{align}
We choose $\lambda < \frac{\epsilon}{12}$, and obtain $\mathrm{Pr}\{E_1\} \rightarrow 0$ as $m \rightarrow \infty$.

Similarly, we can also prove that $\mathrm{Pr}(E_i(k))\rightarrow 0$ for $i=2,3$ and $k=1,2$ as $m,n \rightarrow \infty$ using standard techniques. We can also obtain
\begin{align}
\sum_{\substack{j_1 \neq i_1, \\ j_2 = i_2}} & \mathrm{Pr} \left\{(s_1^m(j_1), s_2^m(j_2), W_k^m) \in  T^{(m)}_{[S_1, S_2, W_k]_{\gamma_{k}}} \right\} \mathrm{Pr} \left\{ (x_1^n(j_1), x_2^n(j_2), Y_k^n) \in  T^{(n)}_{[X_1, X_2, Y_k]_{\delta_k}} \right\} \nonumber \\
&\leq 2^{m[H(S_1)+\frac{\epsilon}{2}] - m[I(S_1;S_2, W_k)-\lambda] - n[I(X_1;Y_k|X_2)-\lambda]} \label{ineq_type1}\\
&= 2^{-m[H(S_1|S_2,W_k) - b I(X_1;Y_k|X_2)- (b+1) \lambda - \frac{\epsilon}{2}]} \nonumber\\
&= 2^{-m[ \frac{\epsilon}{2} - (b+1) \lambda]} \label{ineq_type1b}
\end{align}
where in (\ref{ineq_type1}) we used (\ref{type0}) and (\ref{type1}); and (\ref{ineq_type1b}) holds if the conditions in the theorem hold.

A similar bound can be found for the second summation in (\ref{err_prob}). For the third one, we have the following bound.
\begin{align}
\sum_{\substack{j_1 \neq i_1, \\ j_2 \neq i_2}} & \mathrm{Pr} \left\{(s_1^m(j_1), s_2^m(j_2), W_k^m) \in  T^{(m)}_{[S_1, S_2, W_k]_{\gamma_{k}}}\right\} \mathrm{Pr} \left\{ (x_1^n(j_1), x_2^n(j_2), Y_k^n) \in  T^{(n)}_{[X_1, X_2, Y]_{\delta_k}} \right\} \nonumber \\
& \leq 2^{m[H(S_1)+\epsilon/2] + m[H(S_2)+\epsilon/2]}  2^{-m[I(S_1;S_2,W_k) + I(S_2;S_1,W_k)-I(S_1;S_2|W_k)]-\lambda]}  2^{- n[I(X_1, X_2;Y_k)-\lambda]} \label{ineq_type2} \\
&\leq 2^{-m[H(S_1|S_2,W_k)+ H(S_2|S_1,W_k) - b I(X_1, X_2;Y_k)- (b+1) \lambda -\epsilon ]} \nonumber\\
&= 2^{-m[ \epsilon - (b+1) \lambda]}, \label{ineq_type2b}
\end{align}
where  (\ref{ineq_type2}) follows from (\ref{type0}) and (\ref{type2}); and (\ref{ineq_type2b}) holds if the conditions in the theorem hold.

Choosing $\lambda < \min \left\{\frac{\epsilon}{12}, \frac{\epsilon}{2(b+1)} \right\}$, we can make sure that
all terms of the summation in (\ref{err_prob}) also vanish as $m,n \rightarrow \infty$. Any
rate pair in the convex hull can be achieved by time sharing, hence
the time-sharing random variable $Q$. The cardinality bound on $Q$ follows from the classical arguments.
\end{proof}

\begin{figure*}
\centering
\psfrag{S1}{$S_1^m$} \psfrag{S2}{$S_2^m$}
\psfrag{X1}{$X_1^n$} \psfrag{X2}{$X_2^n$}
\psfrag{Y1}{$Y_1^n$} \psfrag{Y2}{$Y_2^n$}
\psfrag{W1}{$W_1^m$} \psfrag{W2}{$W_2^m$}
\psfrag{hS1}{$(\hat{S}_{1,1}^m, \hat{S}_{1,2}^m)$} \psfrag{hS2}{$(\hat{S}_{2,1}^m, \hat{S}_{2,2}^m)$}
\psfrag{Tx1}{$\mathrm{Tx}_1$}
\psfrag{Tx2}{$\mathrm{Tx}_2$}
\psfrag{Rx1}{$\mathrm{Rx}_1$}
\psfrag{Rx2}{$\mathrm{Rx}_2$}
\psfrag{p12}{$p(s_1, w_2)$}\psfrag{p21}{$p(s_2, w_1)$}
\psfrag{pxy}{$p(y_1, y_2|x_1, x_2)$}
\includegraphics[width=5in]{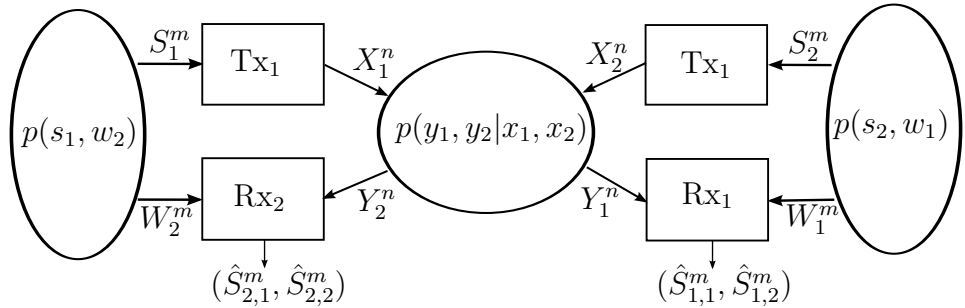}
\caption{Compound multiple access channel in which the transmitter 1 (2) and receiver 2 (1) are located close to each other, and hence have correlated observations, independent of the other pair, i.e., $S_1$ is independent of $(S_2, W_1)$ and $S_2$ is independent of $(S_1, W_2)$ .} \label{f:geog_cMAC}
\end{figure*}

We next prove that the conditions in Theorem \ref{t:ach2} are also
necessary to achieve a source-channel rate of $b$ for some special settings, hence, answering question (2) affirmatively for these cases. We first consider the case in which $S_1$ is independent of $(S_2, W_1)$ and $S_2$ is independent of $(S_1, W_2)$ . This might model a scenario in which $\mathrm{Rx}_1$ ($\mathrm{Rx}_2$) and $\mathrm{Tx}_2$ ($\mathrm{Tx}_1$) are located close to each other, thus having correlated observations, while the two transmitters are far away from each other (see Fig. \ref{f:geog_cMAC}).

\begin{thm}\label{t:c1_comp_mac}
Consider lossless transmission of arbitrarily correlated sources $S_1$ and $S_2$ over a DM compound MAC with side information $W_1$ and $W_2$,
where $S_1$ is independent of $(S_2, W_1)$ and $S_2$ is independent of $(S_1, W_2)$ . Separation (in the operational sense) is optimal for this setup, and the source-channel rate $b$ is achievable if, for $(k,m) \in \{(1,2), (2,1)\}$,
\begin{eqnarray}
H(S_k) < b I(X_k;Y_k |X_m,Q), \label{e:c1_cmac1} \\
H(S_m| W_k) < b I(X_m;Y_k |X_k,Q), \label{e:c1_cmac2}
\end{eqnarray}
and
\begin{eqnarray}
H(S_k) + H(S_m |W_k) < b I(X_k, X_m ;Y_k |Q), \label{e:c1_cmac3}
\end{eqnarray}
for some $|\mathcal{Q}|\leq 4$ and input distribution of the form
\begin{eqnarray}
p(q, x_1, x_2) = p(q)p(x_1|q)p(x_2|q). \label{cmac_c1_pd}
\end{eqnarray}

Conversely, if source-channel rate $b$ is achievable, then (\ref{e:c1_cmac1})-(\ref{e:c1_cmac3}) hold with $<$ replaced by $\leq$ for an input probability distribution of the form given in (\ref{cmac_c1_pd}).
\end{thm}
\begin{proof}
Achievability follows from Theorem \ref{t:ach2}, and the converse proof is given in Appendix \ref{App:cMAC_op_sep}.
\end{proof}

\begin{figure*}
\centering
\psfrag{S1}{$S_1^m$} \psfrag{S2}{$S_2^m$}
\psfrag{X1}{$X_1^n$} \psfrag{X2}{$X_2^n$}
\psfrag{W1}{$W_1^m$} \psfrag{W2}{$W_2^m$}
\psfrag{hS1}{$(\hat{S}_{1,1}^m, \hat{S}_{1,2}^m)$} \psfrag{hS2}{$(\hat{S}_{2,1}^m, \hat{S}_{2,2}^m)$}
\psfrag{Y11}{$Y_{1,1}^m$}\psfrag{Y12}{$Y_{1,2}^m$}
\psfrag{Y21}{$Y_{2,1}^m$}\psfrag{Y22}{$Y_{2,2}^m$}
\psfrag{Tx1}{$\mathrm{Tx}_1$}
\psfrag{Tx2}{$\mathrm{Tx}_2$}
\psfrag{Rx1}{$\mathrm{Rx}_1$}
\psfrag{Rx2}{$\mathrm{Rx}_2$}
\psfrag{p12}{$p(y_{1,1}, y_{1,2} | x_1)$}
\psfrag{p21}{$p(y_{2,1}, y_{2,2} | x_2)$}
\includegraphics[width=5in]{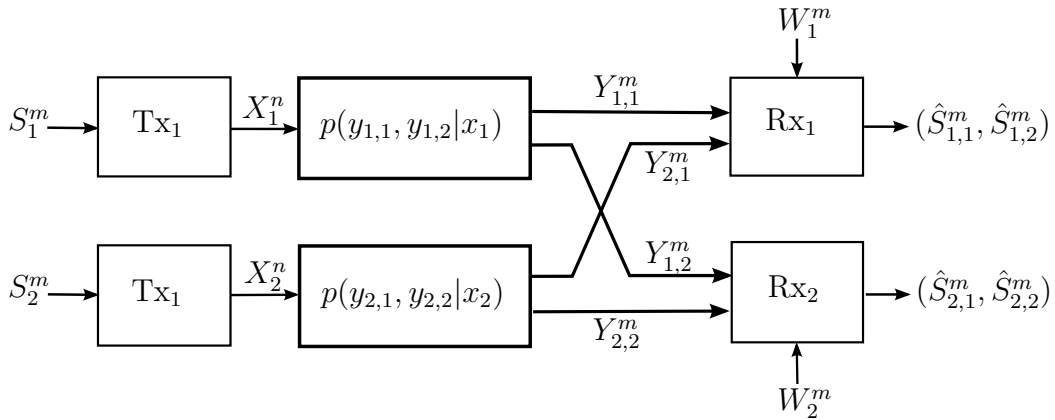}
\caption{Compound multiple access channel with correlated sources and correlated side information with no multiple access interference.} \label{f:ortcMAC}
\end{figure*}

Next, we consider the case in which there is no multiple access interference at the receivers (see Fig. \ref{f:ortcMAC}). We let $Y_k = (Y_{1,k}, Y_{2,k})$ $k=1,2$, where the memoryless channel is characterized by
\begin{eqnarray}
p(y_{1,1}, y_{2,1}, y_{1,2}, y_{2,2} | x_1, x_2) = p(y_{1,1}, y_{1,2}|x_1)p(y_{2,1}, y_{2,2}|x_2). \label{ortcmac_pd}
\end{eqnarray}
On the other hand, we allow arbitrary correlation among the sources and the side information. However, since there is no multiple access interference, using the source correlation to create correlated channel codewords does not enlarge the rate region of the channel. We also remark that this model is not equivalent to two independent broadcast channels with side information. The two encoders interact with each other through the correlation among their sources.

\begin{thm}\label{t:ort_comp_mac}
Consider lossless transmission of arbitrarily correlated sources $S_1$ and $S_2$ over a DM compound MAC with no multiple access interference characterized by (\ref{ortcmac_pd}) and receiver side information $W_1$ and $W_2$ (see Fig. \ref{f:ortcMAC}). Separation (in the operational sense) is optimal for this setup, and the source-channel rate $b$ is achievable if, for $(k,m)=\{(1,2), (2,1)\}$
\begin{eqnarray}
H(S_k| S_m, W_k) &<& b I(X_k;Y_{k,k}), \label{e:ort_cmac1} \\
H(S_m| S_k, W_k) &<& b I(X_m;Y_{m,k}), \label{e:ort_cmac2}
\end{eqnarray}
and
\begin{eqnarray}
H(S_k, S_m| W_k) &<& b [ I(X_k; Y_{k,k}) + I(X_m ;Y_{m,k})], \label{e:ort_cmac3}
\end{eqnarray}
for an input distribution of the form
\begin{eqnarray}
p(q, x_1, x_2) = p(q)p(x_1|q)p(x_2|q). \label{ort_cmac_pd}
\end{eqnarray}

Conversely, if the source-channel rate $b$ is achievable, then (\ref{e:c1_cmac1})-(\ref{e:c1_cmac3}) hold with $<$ replaced by $\leq$ for an input probability distribution of the form given in (\ref{cmac_c1_pd}).
\end{thm}
\begin{proof}
The achievability follows from Theorem \ref{t:ach2} by letting $Q$ be constant and taking into consideration the characteristics of the channel, where $(X_1, Y_{1,1}, Y_{1,2})$ is independent of $(X_2, Y_{2,1}, Y_{2,2})$. The converse can be proven similarly to Theorem \ref{t:c1_comp_mac}, and will be omitted for the sake of brevity.
\end{proof}

Note that the model considered in Theorem \ref{t:ort_comp_mac} is a generalization of the model in \cite{Barros} (which is a special case of the more general network studied in \cite{Han}) to more than one receiver. Theorem \ref{t:ort_comp_mac} considers correlated receiver side information which can be incorporated into the model of \cite{Barros} by considering an additional transmitter sending this side information over an infinite capacity link. In this case, using \cite{Barros}, we observe that informational source-channel separation is optimal. However, Theorem \ref{t:ort_comp_mac} argues that this is no longer true when the number of sink nodes is greater than one even when there is no receiver side information.

The model in Theorem \ref{t:ort_comp_mac} is also considered in \cite{Coleman} in the special case of no side information at the receivers. In the achievability scheme of \cite{Coleman}, transmitters first randomly bin their correlated sources, and then match the bins to channel codewords. Theorem \ref{t:ort_comp_mac} shows that we can achieve the same optimal performance without explicit binning even in the case of correlated receiver side information.

In both Theorem \ref{t:c1_comp_mac} and Theorem \ref{t:ort_comp_mac}, we provide the optimal source-channel matching conditions for lossless transmission. While general matching conditions are not known for compound MACs, the reason we are able to resolve the problem in these two cases is the lack of multiple access interference from users with correlated sources. In the first setup the two sources are independent, hence it is not possible to generate correlated channel inputs, while in the second setup, there is no multiple access interference, and thus there is no need to generate correlated channel inputs. We note here that the optimal source-channel rate in both cases is achieved by operational separation answering both question (2) and question (4) affirmatively. The supoptimality of informational separation in these models follows from \cite{Tuncel}, since the broadcast channel model studied in \cite{Tuncel} is a special case of the compound MAC model we consider. We refer to the example provided in \cite{Coleman} for the suboptimality of informational separation for the setup of Theorem \ref{t:ort_comp_mac} even without side information at the receives.

Finally, we consider the special case in which the two receivers share common side information, i.e., $W_1=W_2=W$, in which case $S_1-W-S_2$ form a Markov chain. For example this models the scenario in which the two receivers are close to each other, hence they have the same side
information. The following theorem proves the optimality of informational separation under these conditions.

\begin{thm}\label{t:comp_mac2}
Consider lossless transmission of correlated sources $S_1$ and $S_2$ over a DM compound MAC with common receiver side information $W_1=W_2=W$ satisfying $S_1-W-S_2$. Separation (in the informational sense) is optimal in this setup, and the source-channel rate $b$ is achievable if, for $k=1$ and $2$,
\begin{eqnarray}
H(S_1|W) &<& b \cdot I(X_1 ; Y_k | X_2, Q), \label{sep_mac_SI_1} \\
H(S_2|W) &<& b \cdot I(X_2 ; Y_k | X_1, Q),  \label{sep_mac_SI_2} \nonumber
\end{eqnarray}
and
\begin{eqnarray}
H(S_1|W) + H(S_2|W) &<& b \cdot I(X_1, X_2 ; Y_k |Q),  \label{sep_mac_SI_3} \nonumber
\end{eqnarray}
for some joint distribution $p(q, x_1, x_2, y) = p(q)p(x_1|q)$ $p(x_2|q)p(y|x_1, x_2)$, with $|\mathcal{Q}| \leq 4$.

Conversely, if the source-channel rate $b$ is achievable, then (\ref{sep_mac_SI_1})-(\ref{sep_mac_SI_3}) hold with $<$ replaced by $\leq$ for an input probability distribution of the form given above.
\end{thm}

\begin{proof}
The achievability follows from informational source-channel separation, i.e, Slepian-Wolf compression conditioned on the receiver side information followed by an optimal compound MAC coding. The proof of the converse follows similarly to the proof of Theorem \ref{t:sep_mac}, and is omitted for brevity.
\end{proof}

\section{Interference channel with correlated sources}\label{s:ic}

In this section, we consider the interference channel (IC) with correlated sources and side information. In the IC each transmitter wishes to communicate only with its corresponding receiver, while the two simultaneous transmissions interfere with each other. Even when the sources and the side information are all independent, the capacity region of the IC is in general not known. The best achievable scheme is given in \cite{Han_Kobayashi}. The capacity region can be characterized in the strong interference case \cite{Sato, Costa:IT:87}, where it coincides with the capacity region of the compound multiple access channel, i.e., it is optimal for the receivers to decode both messages. The interference channel has gained recent interest due to its practical value in cellular and cognitive radio systems. See \cite{Shang:IT:07} - \cite{Annepureddy:IT:07} and references therein for recent results relating to the capacity region of various interference channel scenarios.

For encoders $f_i^{(m,n)}$ and decoders $g_i^{(m,n)}$, the probability of error for the interference channel is given as
\begin{eqnarray}
P_{e}^{(m,n)} &\triangleq& Pr\left\{\bigcup_{k=1,2} S_k^m \neq \hat{S}_{k,k}^m \right\} \nonumber \\
&=& \sum_{(s_1^m,s_2^m) \in \mathcal{S}_1^m \times \mathcal{S}_2^m} p(s_1^m, s_2^m) P\left\{ \bigcup_{k=1,2} \hat{s}_{k,k}^m \neq s_k^m \big| (S_1^m,S_2^m) = (s_1^m,s_2^m)\right\}. \nonumber
\end{eqnarray}
In the case of correlated sources and receiver side information, sufficient conditions for the compound MAC model given in Theorem \ref{t:ach_cMAC} and Theorem \ref{t:ach2} serve as sufficient conditions for the IC as well, since we can constrain both receivers to obtain lossless reconstruction of both sources. Our goal here is to characterize the conditions under which we can provide a converse and achieve either informational or operational separation similar to the results of Section \ref{s:cMAC}. In order to extend the necessary conditions of Theorem \ref{t:c1_comp_mac} and Theorem \ref{t:comp_mac2} to ICs, we will define the `strong source-channel interference' conditions. Note that the interference channel version of Theorem \ref{t:ort_comp_mac} is trivial since the two transmissions do not interfere with each other.

The regular strong interference conditions given in \cite{Sato}
correspond to the case in which, for all input distributions at
transmitter $\mathrm{Tx}_1$, the rate of information flow to receiver $\mathrm{Rx}_2$ is higher
than the information flow to the intended receiver $\mathrm{Rx}_1$. A similar
condition holds for transmitter $\mathrm{Tx}_2$ as well. Hence there is no rate loss if both receivers decode the messages of both transmitters. Consequently, under strong
interference conditions, the capacity region of the IC is equivalent
to the capacity region of the compound MAC. However, in the joint
source-channel coding scenario, the receivers have access to
correlated side information. Thus while calculating the total rate
of information flow to a particular receiver, we should not only
consider the information flow through the channel, but also the
mutual information that already exists between the source and the
receiver side information.

We first focus on the scenario of Theorem \ref{t:c1_comp_mac} in which the source $S_1$ is
independent of $(S_2, W_1)$ and $S_2$ is independent of $(S_2, W_1)$.

\begin{defn}\label{d:strong_int}
For the interference channel in which $S_1$ is independent of $(S_2, W_1)$ and $S_2$ is independent of $(S_2, W_1)$, we say that the \emph{strong source-channel interference conditions} are satisfied for a source-channel rate $b$ if,
\begin{align}
b\cdot I(X_1;Y_1 |X_2) \leq b\cdot I(X_1;Y_2 |X_2) + I(S_1;W_2), \label{st_int1}
\end{align}
and
\begin{align}
b\cdot I(X_2;Y_2 |X_1) \leq b\cdot I(X_2;Y_1 |X_1) + I(S_2;W_1),
\label{st_int2}
\end{align}
for all distributions of the form $p(w_1, w_2, s_1, s_2, x_1, x_2)=
p(w_1, w_2, s_1, s_2)p(x_1|s_1) p(x_2|s_2)$.
\end{defn}

For an IC satisfying these conditions, we next prove the
following theorem.

\begin{thm}\label{t:op_sep_int}
Consider lossless transmission of $S_1$ and $S_2$ over a DM IC with side information $W_1$ and $W_2$, where $S_1$ is independent of $(S_2, W_1)$ and $S_2$ is independent of $(S_2, W_1)$. Assuming that the strong source-channel interference conditions of Definition \ref{d:strong_int} are satisfied for $b$, separation (in the informational sense) is optimal. The source-channel rate $b$ is achievable if, the conditions (\ref{e:cmac1})-(\ref{e:cmac3}) in Theorem \ref{t:ach2} hold. Conversely, if rate $b$ is achievable, then the conditions in Theorem \ref{t:ach2} hold with $<$ replaced by $\leq$.
\end{thm}

Before we proceed with the proof of the theorem, we first prove the following lemma.
\begin{lem}\label{l:str_int}
If $(S_1, W_2)$ is independent of $(S_2, W_1)$ and the strong
source-channel interference conditions
(\ref{st_int1})-(\ref{st_int2}) hold, then we have
\begin{eqnarray}
I(X_2^n;Y_2^n |X_1^n) & \leq&  I(X_2^n;Y_1^n |X_1^n) + I(S_2^m;W_1^m), \label{str_vec1}
\end{eqnarray}
and
\begin{eqnarray}
I(X_1^n;Y_1^n |X_2^n) & \leq&  I(X_1^n;Y_2^n |X_2^n) + I(S_1^m;W_2^m), \label{str_vec2}
\end{eqnarray}
for all $m$ and $n$ satisfying $n/m=b$.
\end{lem}

\begin{proof}
To prove the lemma, we follow the techniques in \cite{Costa:IT:87}. Condition (\ref{st_int2}) implies
\begin{eqnarray}
I(X_2;Y_2 |X_1,U) - I(X_2;Y_1|X_1,U) \leq \frac{1}{b} I(S_2;W_1)
\label{secineq}
\end{eqnarray}
for all $U$ satisfying $U-(X_1,X_2)-(Y_1,Y_2)$.

Then as in \cite{Costa:IT:87}, we can obtain
\begin{align}
I (X_2^n;Y_2^n  |X_1^n) - I(X_2^n;Y_1^n |X_1^n) = & I(X_{2n};Y_{2n}|X_{1}^n, Y_2^{n-1})  - I(X_{2n};Y_{1n} |X_{1}^n, Y_2^{n-1}) \nonumber\\
& + I(X_{2}^{n-1};Y_{2}^{n-1} |X_{1}^{n}, Y_{1n}) - I(X_{2}^{n-1};Y_{1}^{n-1} |X_{1}^n, Y_{1n}) \nonumber\\
= & I(X_{2n};Y_{2n}|X_{1n})  - I(X_{2n};Y_{1n} |X_{1n}) \nonumber \\
& + I(X_{2}^{n-1};Y_{2}^{n-1} |X_{1}^{n-1}) - I(X_{2}^{n-1};Y_{1}^{n-1} |X_{1}^{n-1}) \nonumber\\
= & \sum_{i=1}^n [I(X_{2i};Y_{2i} |X_{1i}) - I(X_{2i};Y_{1i} |X_{1i})].  \nonumber
\end{align}
Using the hypothesis (\ref{st_int2}) of the theorem, we obtain
\begin{align}
I (X_2^n;Y_2^n  |X_1^n) - I(X_2^n;Y_1^n |X_1^n) & \leq  \frac{n}{b}I(S_2;W_1) \nonumber\\
& = I(S_2^m ; W_1^m). \nonumber
\end{align}
Eqn. (\ref{str_vec2}) follows similarly.
\end{proof}

\begin{proof}\textit{(of Theorem \ref{t:op_sep_int})}
Achievability follows by having each receiver decode both $S_1$ and $S_2$, and then using Theorem \ref{t:ach_cMAC}. We next prove the converse. From (\ref{cineq11})-(\ref{cineq17}), we have
\begin{eqnarray}
\frac{1}{n}I(X_1^n;Y_1^n | X_2^n) \geq \frac{1}{b} \left[ H(S_1)- \delta(P_e^{(m,n)}) \right]. \label{ieq0}
\end{eqnarray}

We can also obtain
\begin{align}
\frac{1}{n}I(X_1^n; Y_2^n | X_2^n) &\geq \frac{1}{n} [I(X_1^n;Y_1^n |X_2^n) - I(S_1^m;W_2^m)], \label{ieq1} \\
        &= \frac{1}{b} [H(S_1) -\delta(P_e^{(m,n)})] - \frac{1}{n} I(S_1^m;W_2^m), \label{ieq2} \\
        &= \frac{1}{b} [H(S_1|W_2) - \delta(P_e^{(m,n)})],   \label{ieq3}
\end{align}
in which (\ref{ieq1}) follows from (\ref{str_vec2}), and (\ref{ieq2}) from (\ref{ieq0}).

Finally for the joint mutual information, we have
\begin{align}
\frac{1}{n}I(X_1^n, X_2^n;Y_1^n) = & \frac{1}{n}[I(X_1^n; Y_1^n) + I(X_2^n;Y_1^n |X_1^n)], \nonumber \\
\geq & \frac{1}{n}[I(S_1^m; Y_1^n) + I(X_2^n;Y_2^n |X_1^n) - I(S_2^m ; W_1^m)],  \label{sineq22} \\
\geq & \frac{1}{n}[I(S_1^m; Y_1^n) + I(S_2^m;Y_2^n |X_1^n) - I(S_2^m ; W_1^m)],  \label{sineq23} \\
= & \frac{1}{n}[H(S_1^m) - H(S_1^m | Y_1^n) + H(S_2^m|X_1^n) - H(S_2^m|Y_2^n,X_1^n)  \nonumber\\
&~~~~~ + H(S_2^m |W_1^m) - H(S_2^m)], \nonumber \\
\geq & \frac{1}{n}[H(S_1^m) - H(S_1^m | Y_1^n) - H(S_2^m|Y_2^n) + H(S_2^m |W_1^m) ],  \label{sineq24} \\
= & \frac{1}{n}[H(S_1^m) - H(S_1^m | Y_1^n, W_1^m) - H(S_2^m|Y_2^n, W_2^m) + H(S_2^m |W_1^m) ],  \label{sineq25} \\
\geq & \frac{1}{b}[H(S_1) + H(S_2 |W_1) - 2\delta(P_e^{(m,n)}) ], \label{sineq26}
\end{align}
for any $\epsilon>0$ and large enough $m$ and $n$, where $(\ref{sineq22})$ follows from the data processing inequality and (\ref{str_vec1}); $(\ref{sineq23})$ follows from the data processing inequality since $S_2^m-X_2^n-Y_2^n$ form a Markov chain given $X_1^n$; $(\ref{sineq24})$ follows from the independence of $X_1^n$ and $S_2^m$ and the fact that conditioning reduces entropy; $(\ref{sineq25})$ follows from the fact that $S_1$ is independent of $(S_2, W_1)$ and $S_2$ is independent of $(S_2, W_1)$; and $(\ref{sineq26})$ follows from Fano's inequality. The rest of the proof closely resembles that of Theorem \ref{t:c1_comp_mac}.
\end{proof}

Next, we consider the IC version of the case in Theorem \ref{t:comp_mac2}, in which the
two receivers have access to the same side information $W$ and with this side information the sources are independent. While we still have
correlation between the sources and the common receiver side
information, the amount of mutual information arising from this
correlation is equivalent at both receivers since $W_1 = W_2$. This
suggests that the usual strong interference channel conditions suffice
to obtain the converse result. We have the following theorem for
this case.

\begin{thm}
Consider lossless transmission of correlated sources $S_1$ and $S_2$ over
the strong IC with common receiver side information $W_1=W_2=W$
satisfying $S_1-W-S_2$. Separation (in the informational sense) is optimal in this setup, and the source-channel rate $b$ is achievable if and only if the
conditions in Theorem \ref{t:comp_mac2} hold.
\end{thm}

\begin{proof}
The proof follows from arguments similar to those in the proof of Theorem \ref{t:comp_mac2} and results in \cite{Gunduz:DCC:07}, where we incorporate the strong interference conditions.
\end{proof}

\section{Two-way channel with correlated sources}\label{s:twc}

\begin{figure*}
\centering
\psfrag{S1}{$S_1^m$} \psfrag{S2}{$S_2^m$}
\psfrag{X1}{$X_1^n$} \psfrag{X2}{$X_2^n$}
\psfrag{Y1}{$Y_1^n$} \psfrag{Y2}{$Y_2^n$}
\psfrag{hS1}{$\hat{S}_{1}^m$} \psfrag{hS2}{$\hat{S}_2^m$}
\psfrag{pxy}{$p(y_1,y_2|x_1,x_2)$}
\psfrag{U1}{User 1} \psfrag{U2}{User 2}
\psfrag{Two}{Two-way} \psfrag{Ch}{Channel}
\includegraphics[width=6in]{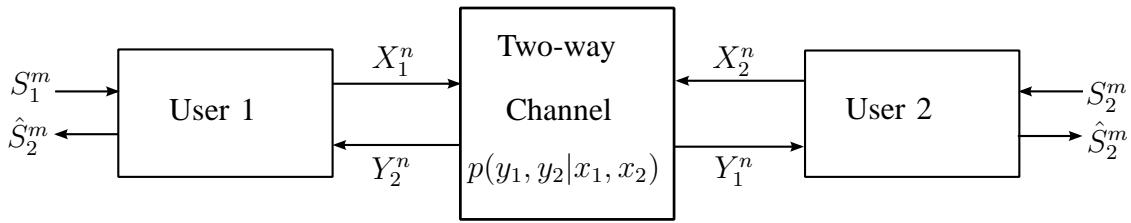}
\caption{The two-way channel model with correlated sources.} \label{f:twc}
\end{figure*}

In this section, we consider the two-way channel scenario with correlated source sequences (see Fig. \ref{f:twc}). The two-way channel model was introduced by Shannon \cite{Shannon2} who gave inner and outer bounds on the capacity region. Shannon showed that his inner bound is indeed the capacity region of the ``restricted'' two-way channel, in which the channel inputs of the users depend only on the messages (not on the previous channel outputs). Several improved outer bounds are given in \cite{Zhang:IT:86}-\cite{Tandon:Asil:07} using the ``dependence-balance bounds'' proposed by Hekstra and Willems.

In \cite{Shannon2} Shannon also considered the case of correlated sources, and showed by an example that by exploiting the correlation structure of the sources we might achieve rate pairs given by the outer bound. Here we consider arbitrarily correlated sources and provide an achievability result using the coding scheme for the compound MAC model in Section \ref{s:cMAC}. It is possible to extend the results to the scenario where each user also has side information correlated with the sources.

In the general two-way channel model, the encoders observe the past channel outputs and hence they can use these observations for encoding future channel input symbols. The encoding function at user $i$ at time instant $j$ is given by
\begin{eqnarray}
f_{i,j}: \mathcal{S}_i^m \times \mathcal{Y}_i^{j-1} \rightarrow \mathcal{X}_{i},
\end{eqnarray}
for $i=1,2$. The probability of error for the two-way channel is given as
\begin{eqnarray}
P_{e}^{(m,n)} &\triangleq& Pr\left\{\bigcup_{k=1,2} S_k^m \neq \hat{S}_k^m \right\} \nonumber \\
&=& \sum_{(s_1^m,s_2^m) \in \mathcal{S}_1^m \times \mathcal{S}_2^m} p(s_1^m, s_2^m) P\left\{ \bigcup_{k=1,2} \hat{s}_k^m \neq s_k^m \big| (S_1^m,S_2^m) = (s_1^m,s_2^m)\right\}. \nonumber
\end{eqnarray}
Note that, if we only consider restricted encoders at the users, than the system model is equivalent to the compound MAC model of Fig. \ref{f:cMAC} with $W_1^m=S_1^m$ and $W_2^m=S_2^m$. From Theorem \ref{t:ach_cMAC} we obtain the following corollary.

\begin{cor}\label{cor:twc:ach}
In lossless transmission of arbitrarily correlated sources $(S_1, S_2)$ over a DM two-way channel, the source-channel rate $b=1$ is achievable if
\begin{eqnarray}
H(S_1|S_2) &<& I(X_1 ; Y_2 | X_2, S_2, Q) \mbox{ and } \nonumber \\
H(S_2|S_1) &<& I(X_2 ; Y_1 | X_1, S_1, Q), \nonumber
\end{eqnarray}
for some joint distribution of the form
\[p(q, s_1, s_2, x_1, x_2, y_1, y_2) = p(q) p(s_1, s_2)p(x_1| q,s_1)p(x_2|q, s_2)p(y_1,y_2|x_1,x_2).\]
\end{cor}

Note that here we use the source correlation rather than the correlation that can be created through the inherent feedback available in the two-way channel. This correlation among the channel codewords potentially helps us achieve source-channel rates that cannot be achieved by independent inputs. Shannon's outer bound can also be extended to the case of correlated sources to obtain a lower bound on the achievable source-channel rate as follows.

\begin{prop}\label{cor:twc:Sb}
In lossless transmission of arbitrarily correlated sources $(S_1, S_2)$ over a DM two-way channel, if the source-channel rate $b$ is achievable, then
\begin{eqnarray}
H(S_1|S_2) &<& b I(X_1 ; Y_2 | X_2) \mbox{ and } \nonumber \\
H(S_2|S_1) &<& b I(X_2 ; Y_1 | X_1), \nonumber
\end{eqnarray}
for some joint distribution of the form
\[p(s_1, s_2, x_1, x_2, y_1, y_2) = p(s_1, s_2)p(x_1, x_2)p(y_1,y_2|x_1,x_2).\]
\end{prop}

\begin{proof}
We have
\begin{align}
H(S_1^m | S_2^m) & = I(S_1^m;Y_2^n|S_2^m) + H(S_1^m|S_2^m, Y_2^n) \label{tw:e:0} \\
        & \leq  I(S_1^m;Y_2^n|S_2^m) + m \delta(P_e^{(m,n)}) \label{tw:e:1} \\
        & = H(Y_2^n | S_2^m) - H(Y_2^n | S_1^m, S_2^m) + m \delta(P_e^{(m,n)}) \label{tw:e:2} \\
        & = \sum_{i=1}^n H(Y_{2i} | S_2^m, Y_2^{i-1}) - H(Y_{2i} | S_1^m, S_2^m, Y_2^{i-1}) + m \delta(P_e^{(m,n)}) \label{tw:e:3} \\
        & \leq \sum_{i=1}^n H(Y_{2i} | S_2^m, Y_2^{i-1}, X_2^i) - H(Y_{2i} | S_1^m, S_2^m, Y_2^{i-1}, Y_1^{i-1}, X_{2i}) + m \delta(P_e^{(m,n)}) \label{tw:e:4} \\
        & \leq \sum_{i=1}^n H(Y_{2i} | X_{2i}) - H(Y_{2i} | S_1^m, S_2^m, Y_2^{i-1}, Y_1^{i-1}, X_{1i}, X_{2i}) + m \delta(P_e^{(m,n)}) \label{tw:e:5} \\
        & \leq \sum_{i=1}^n H(Y_{2i} | X_{2i}) - H(Y_{2i} | X_{1i}, X_{2i}) + m\delta(P_e^{(m,n)}) \label{tw:e:6} \\
        & \leq \sum_{i=1}^n I(X_{1i}; Y_{2i} | X_{2i}) + m\delta(P_e^{(m,n)}) \label{tw:e:7}
\end{align}
where (\ref{tw:e:1}) follows from Fano's inequality; (\ref{tw:e:4}) follows since $X_2^k$ is a deterministic function of $(S_2^m, Y_2^{i-1})$ and the fact that conditioning reduces entropy; (\ref{tw:e:5}) follows similarly as $X_1^k$ is a deterministic function of $(S_1^m, Y_1^{i-1})$ and the fact that conditioning reduces entropy; and (\ref{tw:e:6}) follows since $Y_{2i} - (X_{1i}, X_{2i}) - (S_1^m, S_2^m, Y_2^{i-1}, Y_1^{i-1})$ form a Markov chain.

Similarly, we can show that
\begin{align}
H(S_2^m | S_1^m) & \leq \sum_{i=1}^n I(X_{2i}; Y_{1i} | X_{1i}) + m\delta(P_e^{(m,n)}). \label{tw:e:9}
\end{align}

From convexity arguments and letting $m,n \rightarrow \infty$, we obtain
\begin{eqnarray}
H(S_1|S_2) \leq b I(X_1; Y_2|X_2), \label{twc_rate1} \\
H(S_2|S_1) \leq b I(X_2; Y_1|X_1), \label{twc_rate2}
\end{eqnarray}
for some joint distribution $p(x_1, x_2)$.
\end{proof}

\begin{rem}
Note that the lower bound of Proposition \ref{cor:twc:Sb} allows all possible joint distributions for the channel inputs. This lets us express the lower bound in a separable form, since the source correlation becomes useless to introduce any additional structure to the transmitted channel codewords. In general, not all joint channel input distributions can be achieved at the two users, and tighter bounds can be obtained by limiting the set of possible joint distributions as in \cite{Zhang:IT:86}-\cite{Tandon:Asil:07}.

However, if the existing source correlation allows the users to generate the optimal joint channel input distribution, then the achievable region given in Corollary \ref{cor:twc:ach} might meet the upper bound without the need to exploit the feedback to generate further correlation. This has been illustrated by an example in \cite{Shannon2}. Shannon considered correlated binary sources $S_1$ and $S_2$ such that \[P_{S_1S_2}(S_1=0, S_2=1) = P_{S_1S_2}(S_1=1, S_2=0) = 0.275\] and \[P_{S_1S_2}(S_1=1, S_2=1) =0.45,\] and a binary multiplier two-way channel, in which \[\mathcal{X}_1 =\mathcal{X}_2 =\mathcal{Y}_1 =\mathcal{Y}_2 = \{0,1\}\] and \[Y_1=Y_2=X_1 \cdot X_2.\] Using Proposition \ref{cor:twc:Sb}, we can set a lower bound of $b=1$ on the achievable source-channel rate. On the other hand, the source-channel rate of $1$ can be achieved simply by uncoded transmission. Hence, in this example, the correlated source structure enables the transmitter to achieve the optimal joint distribution for the channel inputs without exploiting the inherent feedback in the two-way channel. Note that the Shannon outer bound is not achievable in the case of independent sources in a binary multiplier two-way channel \cite{Zhang:IT:86}, and the achievable rates can be improved by using channel inputs dependent on the previous channel outputs.
\end{rem}

\section{Conclusions}\label{s:conc}

We have considered source and channel coding over multiuser channels with correlated receiver side information. Due to the lack of a general source-channel separation theorem for multiuser channels, optimal performance in general requires joint source-channel coding. Given the difficulty of finding the optimal source-channel rate in a general setting, we have analyzed some fundamental building-blocks of the general setting in terms of separation optimality. Specifically, we have characterized the necessary and sufficient conditions for lossless transmission over various fundamental multiuser channels, such as multiple access, compound multiple access, interference and two-way channels for certain source-channel distributions and structures. In particular, we have considered transmitting correlated sources over the MAC with receiver side information given which the sources are independent, and transmitting independent sources over the MAC with receiver side information given which the sources are correlated. For the compound MAC, we have provided an achievability result, which has been shown to be tight i) when each source is independent of the other source and one of the side information sequences, ii) when the sources and the side information are arbitrarily correlated but there is no multiple access interference at the receivers, iii) when the sources are correlated and the receivers have access to the same side information given which the two sources are independent. We have then showed that for cases (i) and (iii), the conditions provided for the compound MAC are also necessary for interference channels under some strong source-channel conditions. We have also provided a lower bound on the achievable source-channel rate for the two-way channel.

For the cases analyzed in this paper, we have proven the optimality of designing source and channel codes that are statistically independent of each other, hence resulting in a modular system design without losing the end-to-end optimality. We have shown that, in some scenarios, this modularity can be different from the classical Shannon type separation, called the `informational separation', in which comparison of the source coding rate region and the channel capacity region provides the necessary and sufficient conditions for the achievability of a source-channel rate. In other words, informational separation requires the separate codes used at the source and the channel coders to be the optimal source and the channel codes, respectively, for the underlying model. However, following \cite{Tuncel}, we have shown here for a number of multiuser systems that a more general notion of `operational separation' can hold even in cases for which informational separation fails to achieve the optimal source-channel rate. Operational separation requires statistically independent source and channel codes which are not necessarily the optimal codes for the underlying sources or the channel. In the case of operational separation, comparison of two rate regions (not necessarily the compression rate and the capacity regions) that depend only on the source and channel distributions, respectively, provides the necessary and sufficient conditions for lossless transmission of the sources. These results help us to obtain insights into source and channel coding for larger multiuser networks, and potentially would lead to improved design principles for practical implementations.

\appendices

\section{Proof of Theorem \ref{t:sep_mac_vc}}\label{App:sep_mac_vc}

\begin{proof}
The achievability again follows from separate source and channel coding. We first use Slepian-Wolf compression of the sources conditioned on the receiver side information, then transmit the compressed messages using an optimal multiple access channel code.

An alternative approach for the achievability is possible by considering $W_1$ as the output of a parallel channel from $S_1, S_2$ to the receiver. Note that this parallel channel is used $m$ times for $n$ uses of the main channel. The achievable rates are then obtained following the arguments for the standard MAC:
\begin{eqnarray}
m H(S_1) &<& I(S_1^m, X_1^n ; Y_1^n, W_1^m | X_2^n, S_2^m, Q)\\
    &=& I(S_1^m ; W_1^m | S_2^m) + I(X_1^n ; Y_1^n | X_2^n, Q) \\
    &=& m I(S_1;W_1|S_2) + n I(X_1; Y_1|X_2,Q),
\end{eqnarray}
and using the fact that $p(s_1,s_2,w_1) = p(s_1)p(s_2)p(w_1|s_1,s_2)$ we obtain (\ref{e:MAC1_vc}) (similarly for (\ref{e:MAC2_vc}) and (\ref{e:MAC3_vc})). Note that, this approach provides achievable source-channel rates for general joint distributions of $S_1, S_2$ and $W_1$.

For the converse, we use Fano's inequality given in (\ref{fano1}) and (\ref{fano2}). We have
\begin{eqnarray}
\frac{1}{n}I(X_1^n;Y_1^n|X_2^n) &\geq & \frac{1}{n} I(S_1^m ;Y_1^n| X_2^n), \label{vcineq12} \\
&= & \frac{1}{n} I(S_1^m, W_1^m; Y_1^n | X_2^n),  \label{vcineq14} \\
&\geq & \frac{1}{n} I(S_1^m; Y_1^n | X_2^n, W_1^m), \nonumber \\
&\geq& \frac{1}{n} [H(S_1^m| S_2^m, W_1^m)- m \delta (P_e^{(m,n)})], \label{vcineq17}\\
&\geq& \frac{1}{b} [H(S_1| S_2, W_1)- \delta (P_e^{(m,n)})], \nonumber
\end{eqnarray}
where $(\ref{vcineq12})$ follows from the Markov relation
$S_1^m-X_1^n-Y_1^n$ given $X_2^n$; $(\ref{vcineq14})$ from the
Markov relation $W_1^m-(X_2^n, S_1^m)- Y_1^n$; and $(\ref{vcineq17})$ from Fano's inequality (\ref{fano2}).

We also have
\begin{eqnarray}
\frac{1}{n} \sum_{i=1}^n I(X_{1i}; Y_{1,i} | X_{2i}) &\geq& \frac{1}{n} I(X_1^n, X_2^n; Y_1^n)  \nonumber \\
&\geq & \frac{1}{b} [H(S_1| S_2, W_1)- \delta (P_e^{(m,n)})]. \nonumber
\end{eqnarray}
Similarly, we have
\begin{eqnarray}
\frac{1}{n} \sum_{i=1}^n I(X_{2i} ; Y_{1,i} | X_{1i}) &\geq & \frac{1}{b} [H(S_2| S_1, W_1)- \delta (P_e^{(m,n)})], \nonumber
\end{eqnarray}
and
\begin{eqnarray}
\frac{1}{n} \sum_{i=1}^n I(X_{1i},X_{2i} ; Y_{1,i}) &\geq & \frac{1}{b} [H(S_1, S_2 | W_1)- \delta (P_e^{(m,n)})]. \nonumber
\end{eqnarray}
As usual, we let $P_e^{(m,n)} \rightarrow 0$, and introduce the time sharing random variable $Q$ uniformly distributed over $\{1,2,\ldots,n\}$ and independent of all the other random variables. Then we define $X_1\triangleq X_{1Q}$, $X_2\triangleq X_{2Q}$ and $Y_1\triangleq Y_{1Q}$. Note that $Pr\{X_1=x_1, X_2=x_2 |Q = q\} = Pr\{X_1|Q=q \} \cdot Pr\{X_2|Q=q\}$ since the two sources, and hence the channel codewords, are independent of each other conditioned on $Q$. Thus, we obtain (\ref{e:MAC1_vc})-(\ref{e:MAC3_vc}) for a joint distribution of the form (\ref{pd_MAC_v2}).
\end{proof}

\section{Proof of Theorem \ref{t:c1_comp_mac}}\label{App:cMAC_op_sep}

We have
\begin{align} \label{cineq11}
\frac{1}{n}I(X_1^n;Y_1^n | X_2^n) \geq & \frac{1}{n} I(S_1^m;Y_1^n | X_2^n), \\
= &\frac{1}{n} [H(S_1^m| X_2^n)- H(S_1^m|Y_1^n, X_2^n)],  \label{cineq15} \\
\geq &\frac{1}{n} [H(S_1^m)- H(S_1^m|Y_1^n)],  \label{cineq16} \\
\geq &\frac{1}{b} \left[ H(S_1)- \delta(P_e^{(m,n)}) \right],  \label{cineq17}
\end{align}
for any $\epsilon>0$ and sufficiently large $m$ and $n$, where $(\ref{cineq11})$
follows from the conditional data processing inequality since
$S_1^m-X_1^n-Y_1^n$ forms a Markov chain given $X_2^n$;
$(\ref{cineq16})$ from the independence of $S_1^m$ and $X_2^n$ and
the fact that conditioning reduces entropy; and $(\ref{cineq17})$ from
the memoryless source assumption, and from Fano's inequality.

For the joint mutual information, we can write the following set of inequalities:
\begin{align}
& \frac{1}{n}I(X_1^n, X_2^n;Y_1^n) \geq \frac{1}{n} I(S_1^m, S_2^m;Y_1^n), \label{cineq21} \\
        &= \frac{1}{n} I(S_1^m,S_2^m, W_1^m ;Y_1^n), \label{cineq22} \\
        &\geq \frac{1}{n} I(S_1^m,S_2^m ;Y_1^n | W_1^m), \label{cineq23} \\
        &= \frac{1}{n} [H(S_1^m,S_2^m| W_1^m)- H(S_1^m,S_2^m|Y_1^n, W_1^m)],  \nonumber \\
        &= \frac{1}{n} [H(S_1^m) + H(S_2^m| W_1^m) - H(S_1^m,S_2^m|Y_1^n, W_1^m)], \label{cineq26} \\
        &\geq \frac{1}{b} \bigg[ H(S_1) + H(S_2 | W_1) - \delta(P_e^{(m,n)}) \bigg],   \label{cineq27}
\end{align}
for any $\epsilon>0$ and sufficiently large $m$ and $n$, where $(\ref{cineq21})$
follows from the data processing inequality since $(S_1^m, S_2^m) -
(X_1^n, X_2^n) -Y_1^n$ form a Markov chain; $(\ref{cineq22})$ from
the Markov relation $W_1^m-(S_1^m, S_2^m) -Y_1^n$; $(\ref{cineq23})$
from the chain rule and the non-negativity of the mutual
information; $(\ref{cineq26})$ from the independence of $S_1^m$ and
$(S_2^m, W_1^m)$; and $(\ref{cineq27})$ from the memoryless source
assumption and Fano's inequality.

It is also possible to show that
\begin{align}
 \sum_{i=1}^n I(X_{1i}; Y_{1i}|X_{2i}) \geq I(X_1^n; Y_1^n | X_2^n),
\end{align}
and similarly for other mutual information terms. Then, using the
above set of inequalities and letting $P_e^{(m,n)} \rightarrow 0$, we
obtain
\begin{eqnarray}
\frac{1}{b} H(S_1) &\leq& \frac{1}{n} \sum_{i=1}^n I(X_{1i}; Y_{1i}|X_{2i}), \nonumber \\
\frac{1}{b} H(S_2|W_1) &\leq& \frac{1}{n} \sum_{i=1}^n I(X_{2i}; Y_{1i}|X_{1i}), \nonumber
\end{eqnarray}
and
\begin{eqnarray}
\frac{1}{b} (H(S_1) + H(S_2| W_1) ) &\leq& \frac{1}{n} \sum_{i=1}^n
I(X_{1i}, X_{2i}; Y_{1i}), \nonumber
\end{eqnarray}
for any product distribution on $\mathcal{X}_1 \times
\mathcal{X}_2$. We can write similar expressions for the second
receiver as well. Then the necessity of the conditions of Theorem
\ref{t:ach2} can be argued simply by inserting the time-sharing
random variable $Q$.


\end{document}